\documentclass[journal]{IEEEtran}

\usepackage{cite}

\usepackage{graphicx}
\usepackage{xcolor}
\usepackage{tikz}
\usetikzlibrary{arrows,positioning,shapes.geometric,calc,spy,patterns,backgrounds}
\usetikzlibrary{decorations.markings}
\usepackage{pgfplots}
\usepackage{multirow}
\usepackage{upgreek}
\usepackage{amssymb}
\usepackage{amsmath}
\usepackage{cases}
\usepackage{pifont}
\usepackage{bm}
\usepackage{amsthm}

\usepackage{tabularx}
\usepackage{booktabs}
\usepackage{filecontents}
\usepackage{subcaption}
\usepackage[normalem]{ulem}
\newcolumntype{Y}{>{\centering\arraybackslash}X}

\usepackage{mathtools}

\usepackage[]{algorithm2e}
\SetAlCapSkip{1em}

\usepackage{array}
\usepackage{soul}
\definecolor{Paired-2}{RGB}{166,206,227}
\definecolor{Paired-1}{RGB}{31,120,180}
\definecolor{Paired-4}{RGB}{178,223,138}
\definecolor{Paired-3}{RGB}{51,160,44}
\definecolor{Paired-6}{RGB}{251,154,153}
\definecolor{Paired-5}{RGB}{227,26,28}
\definecolor{Paired-8}{RGB}{253,191,111}
\definecolor{Paired-7}{RGB}{255,127,0}
\definecolor{Paired-10}{RGB}{202,178,214}
\definecolor{Paired-9}{RGB}{106,61,154}
\definecolor{Paired-12}{RGB}{255,255,153}
\definecolor{Paired-11}{RGB}{177,89,40}

\usepackage[symbol*]{footmisc}
\DefineFNsymbolsTM{myfnsymbols}{
  \textasteriskcentered *
  \textdagger    \dagger
  \textdaggerdbl \ddagger
  \textsection   \mathsection
  \textbardbl    \|%
  \textparagraph \mathparagraph
}%
\setfnsymbol{myfnsymbols}

\definecolor{Set1-1}{RGB}{228,26,28}
\definecolor{Set1-2}{RGB}{55,126,184}

\begin{document}

\title{Operation Merging for Hardware Implementations of Fast Polar Decoders}

\author{Furkan~Ercan,
		Thibaud~Tonnellier,
        Carlo~Condo,
        and Warren~J.~Gross%
\thanks{F.~Ercan, T.~Tonnellier, C.~Condo and W.~J.~Gross are with the Department of Electrical and Computer Engineering, McGill University, Montr\'eal, Qu\'ebec, Canada. e-mail: furkan.ercan@mail.mcgill.ca, thibaud.tonnellier@mail.mcgill.ca, carlo.condo@mail.mcgill.ca,  warren.gross @mcgill.ca. This article is published in Journal of Signal Processing Systems (JSPS), vol. 91, pp.995-1007 on November 3, 2018. DOI:10.1007/s11265-018-1413-4}}

\maketitle

\begin{abstract}
Polar codes are a class of linear block codes that provably achieves channel capacity. They have been selected as a coding scheme for the control channel of enhanced mobile broadband (eMBB) scenario for $5^{\text{th}}$ generation wireless communication networks (5G) and are being considered for additional use scenarios. As a result, fast decoding techniques for polar codes are essential. Previous works targeting improved throughput for successive-cancellation (SC) decoding of polar codes are semi-parallel implementations that exploit special maximum-likelihood (ML) nodes. In this work, we present a new fast simplified SC (Fast-SSC) decoder architecture. Compared to a baseline Fast-SSC decoder, our solution is able to reduce the memory requirements. We achieve this through a more efficient memory utilization, which also enables to execute multiple operations in a single clock cycle. Finally, we propose new special node merging techniques that improve the throughput further, and detail a new Fast-SSC-based decoder architecture to support merged operations. The proposed decoder reduces the operation sequence requirement by up to $39\%$, which enables to reduce the number of time steps to decode a codeword by $35\%$. ASIC implementation results with 65 nm TSMC technology show that the proposed decoder has a throughput improvement of up to $31\%$ compared to previous Fast-SSC decoder architectures.
\end{abstract}


\IEEEpeerreviewmaketitle

\section{Introduction}
Polar codes, introduced by Ar{\i}kan \cite{arikan09}, are a class of linear block codes that provably achieves channel 
capacity. They have been selected as a coding scheme for enhanced mobile broadband (eMBB) scenario under $5^{\text{th}}$ 
generation wireless communication standards (5G) \cite{38.212,hashemi2017asilomar}, and are also being considered for ultra reliable low-latency 
communication (URLLC) and massive machine-type communication (mMTC) in 5G networks \cite{sharma2017polar,sybis2016channel}. 

Successive cancellation (SC) decoding of polar codes is the original decoding scheme proposed in \cite{arikan09}, and 
can be represented as a binary tree search. However, this approach suffers from long decoding latency due to its 
sequential nature, and mediocre error-correction performance at moderate to short code lengths. In order to reduce the 
latency of SC decoding, SSC \cite{SSC2011} and Fast-SSC \cite{sarkis14} decoders proposed efficient decoding techniques for particular information and frozen bit patterns, called special nodes, 
without affecting the error-correction performance. Compared to conventional SC decoder implementations \cite{TPSC13}, 
Fast-SSC decoding is shown to improve the throughput by an order of magnitude. Further identification and use of special 
nodes were carried out in both SC-based \cite{giard16, giardJSPS, fastssc-sips17} and SC-List based \cite{fastSSCL-TCAS-I,fastSSCL-TSP} 
decoding techniques.

In \cite{sarkis14}, a number of parallel processing elements ($P_e$) allows to achieve high throughput. However, there are two problems regarding the use of parallel processing elements in Fast-SSC decoding. The first issue is that the memory utilization factor of the decoder decreases with increasing $P_e$. Secondly, for nodes with sizes smaller than $P_e$, a single operation is performed where the architecture is able to support multiple operations.

In this work, we present a new Fast-SSC decoder architecture that substantially increases the memory utilization. Unlike the previous Fast-SSC-based 
architectures, the new memory utilization is regardless of the parallelization factor. The new configuration allows an opportunity to perform multiple operations at a 
single step. By observing the distribution of frozen bits, we identify two categories of operation merging scenarios. The first category includes merging branch-type 
operations, where a  leaf node estimation is not included. The second category includes merging of special nodes at the bottom of the SC tree. A subset of them is 
selected for a new SC-based decoder implementation to improve the decoder throughput. Results show that, our proposed decoder reduces the number of operations by up to $
39\%$, which enables to reduce the number of time steps to decode a codeword by $35\%$. Results in 65 nm TSMC CMOS show that the proposed decoder has a throughput improvement of up to $31\%$ compared to previous Fast-SSC decoder architectures, while increasing the memory utilization to $99.6\%$.

This paper is an extension of our previous works in \cite{fastssc-sips17,hashemi2017jetcas}, where memory reduction and operation merging schemes were first introduced, followed by an FGPA implementation. In this paper, we generalize operation merging scenarios and implement a novel decoder architecture with significantly improved throughput.

The rest of this paper is organized as follows: In Section \ref{sec:background}, preliminaries for polar code encoding 
and decoding are reviewed. A new memory design to improve utilization for Fast-SSC decoding is described in Section~\ref{sec:proposed-memory}. Section \ref{sec:proposed-merging} describes operation merging scenarios for Fast-SSC decoding.
In Section \ref{sec:architecture}, a new Fast-SSC decoder architecture is described. ASIC synthesis results for the new 
decoder are presented and compared against state-of-the-art decoder implementations in Section \ref{sec:results}, and
finally concluding remarks are addressed in Section \ref{sec:conclusion}.

\section{Preliminaries}\label{sec:background}

\subsection{Polar Codes}\label{sec:polarcodes}

\begin{figure}
  \centering
  \scalebox{1.00}{
  \begin{tikzpicture}[scale=.65, thick]
  \node [color=darkgray] at (.5,0) {$u_0$} ;
  \node [color=darkgray]at (.5,-1) {$u_1$};
  \node [color=darkgray]at (.5,-2) {$u_2$};
  \node at (.5,-3) {$\boldsymbol{u_3}$};
  \node at (.5,-4) {$\boldsymbol{u_4}$};
  \node at (.5,-5) {$\boldsymbol{u_5}$};
  \node at (.5,-6) {$\boldsymbol{u_6}$};
  \node at (.5,-7) {$\boldsymbol{u_7}$};

  \foreach \x in {-6,-4,-2,0}
  {
    \draw [->] (1,\x) -- (2,\x);
    \draw (1,\x-1) -- (2.25,\x-1);

    \draw (2.25,\x) circle [radius=.25];
    \draw (2,\x) -- (2.5,\x);
    \draw (2.25,\x-.25) -- (2.25,\x+.25);

    \draw [->] (2.25,\x-1) -- (2.25,\x-.25);

    \fill (2.25,\x-1) circle [radius=.1];
  }

  \foreach \x in {-4,0}
  {
    \draw [->] (2.5,\x) -- (5,\x);
    \draw [->] (2.25,\x-1) -- (3.5,\x-1);

    \draw (5.25,\x) circle [radius=.25];
    \draw (5,\x) -- (5.5,\x);
    \draw (5.25,\x-.25) -- (5.25,\x+.25);

    \draw (3.75,\x-1) circle [radius=.25];
    \draw (3.5,\x-1) -- (4,\x-1);
    \draw (3.75,\x-1-.25) -- (3.75,\x-1+.25);

    \draw [->] (2.25,\x-2) -- (5.25,\x-2) -- (5.25,\x-.25);
    \fill (5.25,\x-2) circle [radius=.1];
    \draw [->] (2,\x-3) -- (3.75,\x-3) -- (3.75,\x-1-.25);
    \fill (3.75,\x-3) circle [radius=.1];
  }

  \draw [->] (5.5,0) -- (11,0);
  \draw [->] (4,-1) -- (9.5,-1);
  \draw [->] (5.25,-2) -- (8,-2);
  \draw [->] (3.75,-3) -- (6.5,-3);

  \foreach \x in {-1,0}
  {
    \draw [->] (5.5+1.5*\x,\x-4) -- (11.25+1.5*\x,\x-4) -- (11.25+1.5*\x,\x-.25);
    \draw [->] (5.25+1.5*\x,\x-6) -- (11.25+1.5*\x-3,\x-6) -- (11.25+1.5*\x-3,\x-2-.25);
  }

  \foreach \x in {-3,...,0}
  {
    \draw (11.25+1.5*\x,\x) circle [radius=.25];
    \draw (11+1.5*\x,\x) -- (11.5+1.5*\x,\x);
    \draw (11.25+1.5*\x,\x-.25) -- (11.25+1.5*\x,\x+.25);

    \fill (11.25+1.5*\x,\x-4) circle [radius=.1];

    \draw [->] (11.5+1.5*\x,\x) -- (12.5,\x);
    \draw [->] (11.25+1.5*\x,\x-4) -- (12.5,\x-4);
  }

  \node at (13,0) {$\boldsymbol{x_0}$};
  \node at (13,-1) {$\boldsymbol{x_1}$};
  \node at (13,-2) {$\boldsymbol{x_2}$};
  \node at (13,-3) {$\boldsymbol{x_3}$};
  \node at (13,-4) {$\boldsymbol{x_4}$};
  \node at (13,-5) {$\boldsymbol{x_5}$};
  \node at (13,-6) {$\boldsymbol{x_6}$};
  \node at (13,-7) {$\boldsymbol{x_7}$};

\end{tikzpicture}}
  \caption{Polar code encoding for $PC(8,5)$. Gray indices indicate frozen bits while black indices represent information bits.}
  \label{fig:polarencode}
\end{figure}
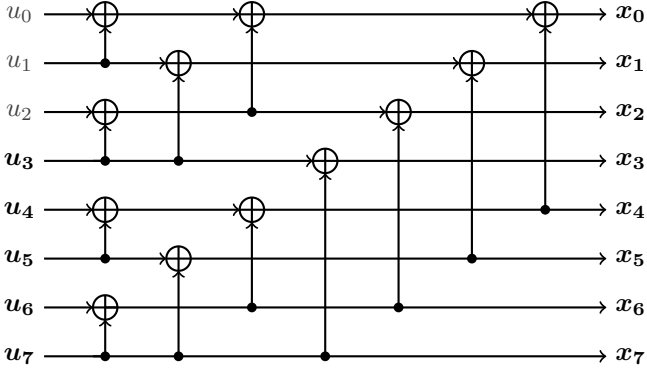

Polar codes are able to achieve channel capacity through channel polarization, that splits $N$ channel utilizations into 
$K$ reliable ones, through which information bits are sent, and $N-K$ unreliable ones, used for frozen bits. A polar code, 
represented as $PC(N,K)$, is a linear block code of length $N = 2^n$ and rate $R = K/N$. Encoding of a polar code can be 
represented by a matrix multiplication:
\begin{equation}\label{eq:enc}
\boldsymbol{x_0^{N-1}} = \boldsymbol{u_0^{N-1}}G^{\otimes n}\text{,}
\end{equation}
where $\boldsymbol{u_0^{N-1}} = \{u_0,u_1,\ldots,u_{N-1}\}$ is the input vector, $\boldsymbol{x_0^{N-1}} = \{x_0,x_1,\ldots,x_{N-1}\}$ 
is the encoded vector, and the generator matrix $G^{\otimes n}$ is the $n$-th Kronecker product of the polar code matrix 
$G = \left[\begin{smallmatrix} 1&0\\ 1&1 \end{smallmatrix} \right]$. A polar code of length $N$ is composed of two 
concatenated polar codes of length $N/2$; Fig. \ref{fig:polarencode} depicts the encoding process for $PC(8,5)$.

\subsection{Successive Cancellation Decoding}\label{sec:scdecoding}

\begin{figure}
  \centering
  \scalebox{1.00}{
  \begin{tikzpicture}[scale=.75]
\usetikzlibrary{backgrounds}

\filldraw[fill=gray!40!white, draw=black] (+0.00,+0.00) circle [radius=.2];

\filldraw[fill=gray!40!white, draw=black] (-3.00,-1.00) circle [radius=.2];
\filldraw[fill=gray!40!white, draw=black] (+3.00,-1.00) circle [radius=.2];

\filldraw[fill=gray!40!white, draw=black] (-4.50,-2.20) circle [radius=.2];
\filldraw[fill=gray!40!white, draw=black] (-1.50,-2.20) circle [radius=.2];
\filldraw[fill=gray!40!white, draw=black] (+1.50,-2.20) circle [radius=.2];
\filldraw[fill=gray!40!white, draw=black] (+4.50,-2.20) circle [radius=.2];

\filldraw[fill=gray!40!white, draw=black] (-5.25,-3.40) circle [radius=.2];
\filldraw[fill=gray!40!white, draw=black] (-3.75,-3.40) circle [radius=.2];
\filldraw[fill=gray!40!white, draw=black] (-2.25,-3.40) circle [radius=.2];
\filldraw[fill=gray!40!white, draw=black] (-0.75,-3.40) circle [radius=.2];
\filldraw[fill=gray!40!white, draw=black] (+0.75,-3.40) circle [radius=.2];
\filldraw[fill=gray!40!white, draw=black] (+2.25,-3.40) circle [radius=.2];
\filldraw[fill=gray!40!white, draw=black] (+3.75,-3.40) circle [radius=.2];
\filldraw[fill=gray!40!white, draw=black] (+5.25,-3.40) circle [radius=.2];

\filldraw[fill=white!40!white, draw=black] (-5.60,-4.60) circle [radius=.2];
\filldraw[fill=white!40!white, draw=black] (-4.90,-4.60) circle [radius=.2];
\filldraw[fill=white!40!white, draw=black] (-4.10,-4.60) circle [radius=.2];
\filldraw[fill=white!40!white, draw=black] (-3.40,-4.60) circle [radius=.2];
\filldraw[fill=white!40!white, draw=black] (-2.60,-4.60) circle [radius=.2];
\filldraw[fill=black!40!black, draw=black] (-1.90,-4.60) circle [radius=.2];
\filldraw[fill=black!40!black, draw=black] (-1.10,-4.60) circle [radius=.2];
\filldraw[fill=black!40!black, draw=black] (-0.40,-4.60) circle [radius=.2];
\filldraw[fill=white!40!white, draw=black] (+0.40,-4.60) circle [radius=.2];
\filldraw[fill=black!40!black, draw=black] (+1.10,-4.60) circle [radius=.2];
\filldraw[fill=black!40!black, draw=black] (+1.90,-4.60) circle [radius=.2];
\filldraw[fill=black!40!black, draw=black] (+2.60,-4.60) circle [radius=.2];
\filldraw[fill=black!40!black, draw=black] (+3.40,-4.60) circle [radius=.2];
\filldraw[fill=black!40!black, draw=black] (+4.10,-4.60) circle [radius=.2];
\filldraw[fill=black!40!black, draw=black] (+4.90,-4.60) circle [radius=.2];
\filldraw[fill=black!40!black, draw=black] (+5.60,-4.60) circle [radius=.2];

\node [color=black] at (-5.60,-5.10) {$\hat{u}_0$};
\node [color=black] at (-4.90,-5.10) {$\hat{u}_1$};
\node [color=black] at (-4.10,-5.10) {$\hat{u}_2$};
\node [color=black] at (-3.40,-5.10) {$\hat{u}_3$};
\node [color=black] at (-2.60,-5.10) {$\hat{u}_4$};
\node [color=black] at (-1.90,-5.10) {$\hat{u}_5$};
\node [color=black] at (-1.10,-5.10) {$\hat{u}_6$};
\node [color=black] at (-0.40,-5.10) {$\hat{u}_7$};
\node [color=black] at (+0.40,-5.10) {$\hat{u}_8$};
\node [color=black] at (+1.10,-5.10) {$\hat{u}_9$};
\node [color=black] at (+1.90,-5.10) {$\hat{u}_{10}$};
\node [color=black] at (+2.60,-5.10) {$\hat{u}_{11}$};
\node [color=black] at (+3.40,-5.10) {$\hat{u}_{12}$};
\node [color=black] at (+4.10,-5.10) {$\hat{u}_{13}$};
\node [color=black] at (+4.90,-5.10) {$\hat{u}_{14}$};
\node [color=black] at (+5.60,-5.10) {$\hat{u}_{15}$};

\begin{scope}[on background layer]
\draw [-] (+0.00,+0.00) -- (-3.00,-1.00);
\draw [-] (+0.00,+0.00) -- (+3.00,-1.00);

\draw [-] (-3.00,-1.00) -- (-4.50,-2.20);
\draw [-] (-3.00,-1.00) -- (-1.50,-2.20);
\draw [-] (+3.00,-1.00) -- (+1.50,-2.20);
\draw [-] (+3.00,-1.00) -- (+4.50,-2.20);

\draw [-] (-4.50,-2.20) -- (-5.25,-3.40);
\draw [-] (-4.50,-2.20) -- (-3.75,-3.40);
\draw [-] (-1.50,-2.20) -- (-2.25,-3.40);
\draw [-] (-1.50,-2.20) -- (-0.75,-3.40);
\draw [-] (+1.50,-2.20) -- (+0.75,-3.40);
\draw [-] (+1.50,-2.20) -- (+2.25,-3.40);
\draw [-] (+4.50,-2.20) -- (+3.75,-3.40);
\draw [-] (+4.50,-2.20) -- (+5.25,-3.40);

\draw [-] (-5.25,-3.40) -- (-5.60,-4.60);
\draw [-] (-5.25,-3.40) -- (-4.90,-4.60);
\draw [-] (-3.75,-3.40) -- (-4.10,-4.60);
\draw [-] (-3.75,-3.40) -- (-3.40,-4.60);
\draw [-] (-2.25,-3.40) -- (-2.60,-4.60);
\draw [-] (-2.25,-3.40) -- (-1.90,-4.60);
\draw [-] (-0.75,-3.40) -- (-1.10,-4.60);
\draw [-] (-0.75,-3.40) -- (-0.40,-4.60);
\draw [-] (+0.75,-3.40) -- (+0.40,-4.60);
\draw [-] (+0.75,-3.40) -- (+1.10,-4.60);
\draw [-] (+2.25,-3.40) -- (+1.90,-4.60);
\draw [-] (+2.25,-3.40) -- (+2.60,-4.60);
\draw [-] (+3.75,-3.40) -- (+3.40,-4.60);
\draw [-] (+3.75,-3.40) -- (+4.10,-4.60);
\draw [-] (+5.25,-3.40) -- (+4.90,-4.60);
\draw [-] (+5.25,-3.40) -- (+5.60,-4.60);
\end{scope}



\end{tikzpicture}}
  \caption{Successive cancellation decoder tree for $PC(16,10)$.}
  \label{fig:polartree}
\end{figure}
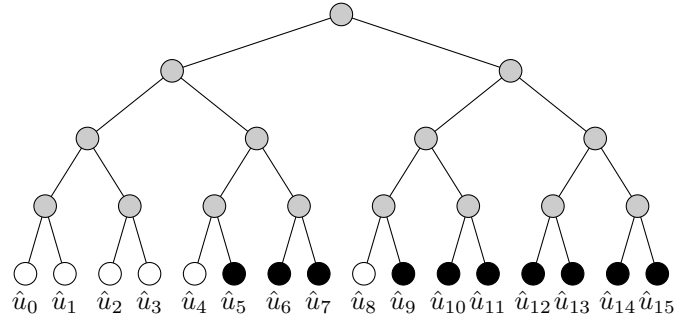

SC decoding \cite{arikan09} can be interpreted as a binary tree search and is explored depth-first, with priority to the left branch.
An example to SC decoder tree for $PC(16,10)$ is shown in Fig.~\ref{fig:polartree}. The root of the tree consists of the 
information obtained by the channel, which is expressed in terms of log-likelihood ratio (LLR) for this work. At each 
stage of the tree, the LLR values $\boldsymbol{\alpha} = \{\alpha_{0},\alpha_{1},..\alpha_{2^S-1}\}$ are passed from a parent node to its child nodes, and hard decision estimates $\boldsymbol{\beta}=\{\beta_{0},\beta_{1},..\beta_{2^S-1}\}$ are passed 
from a child node to its parent node. The soft information passed to left child $\boldsymbol{\alpha^l}$ and right child 
$\boldsymbol{\alpha^r}$ are approximated as
\begin{equation}\label{eqn:alphaleft}
{\alpha}^l_i = \text{~sgn}(\alpha_{i})\text{~sgn}(\alpha_{i+2^{S-1}}) \text{~min}(|\alpha_{i}|,|\alpha_{i+2^{S-1}}|)
\end{equation}
\begin{equation}\label{eqn:alpharight}
{\alpha}^r_i = \alpha_{i+2^{S-1}} + (1-2\beta^{l}_{i})\alpha_{i}
\end{equation}
where $0 \leq i < 2^{S-1}$ for stage $S$, and the root node is at stage $S=\log_2(N)$.

The hard decision estimates, $\boldsymbol{\beta}$ for stage $S$, are calculated via the left and right messages from 
child nodes, $\boldsymbol{\beta^{l}}$ and $\boldsymbol{\beta^{r}}$, as

\begin{equation}\label{eqn:beta}
  \beta_i=\left\{
  \begin{array}{@{}ll@{}}
    \beta^{l}_{i} \oplus \beta^{r}_{i}, & \text{if}~ i \leq 2^{S-1} \\
    \beta^{r}_{i-2^{S-1}}, & \text{otherwise.}
  \end{array}\right.
\end{equation} 

\noindent
where $\boldsymbol{\oplus}$ denotes bitwise XOR operation, and where $0 \leq i < 2^S$. At the leaf nodes, $\beta$ values 
are hard decisions computed by observing the sign bit of their soft information, as

\begin{equation}\label{eqn:beta-leaf}
  \beta_{i}^{leaf}=\left\{
  \begin{array}{@{}ll@{}}
    0, & \text{if}~ \alpha_i^{leaf} \geq 0 \text{ } \text{or } i \in \Phi; \\
    1, & \text{otherwise.}
  \end{array}\right.
\end{equation} 

\noindent
where $i$ represents the node index and $\Phi$ denotes the set of frozen indices.

\subsection{Fast-SSC Decoding}\label{sec:fastssc-old}

Simplified successive cancellation (SSC) decoding \cite{SSC2011} showed that the SC tree can be pruned, avoiding the 
descent in case of nodes whose leaf nodes are either all information bits (Rate-1) or all frozen bits (Rate-0). The Fast-SSC decoding algorithm \cite{sarkis14} evolves SSC 
by identifying special patterns and presenting efficient decoding techniques for such nodes.

\subsubsection{Algorithm}\label{subsec:fastssc_alg}

If we denote an information 
bit with I and a frozen bit with F, in addition to Rate-0 and Rate-1 nodes, special nodes can occur in three other different 
forms in a polar code: repetition (Rep) (FF$\cdot\cdot\cdot$FI), single parity check (SPC) (FI$\cdot\cdot\cdot$II) and a pattern (FFII) which is referred as ML node in \cite{sarkis14}. In this work, we follow the same naming conventions for simplicity. 

A repetition node contains a single information bit; all other nodes are frozen. An information node encoded with frozen 
bits contain the same information bit in all the nodes. The hard decision is made by adding the LLR values together 
and extracting the sign bit of the result:
\begin{equation}\label{eqn:rep}
  {\beta}_i=\left\{
  \begin{array}{@{}ll@{}}
    0, & \text{if}~\sum_{i=0}^{N_v-1} \alpha_i \geq 0 \\
    1, & \text{otherwise.}
  \end{array}\right.
\end{equation} 

In SPC nodes, due to the nature of the polar code construction, the frozen bit represents the parity of all the information bits of the node. Consequently, the parity check for all hard decisions (HDs) (Eq. \ref{eqn:SPC-HD}) of an SPC node must be zero (Eq. \ref{eqn:SPC-parity}).

\begin{equation}\label{eqn:SPC-HD}
  \text{HD}_i=\left\{
  \begin{array}{@{}ll@{}}
    0, & \text{if}~  \alpha_i \geq 0 \\
    1, & \text{otherwise.}
  \end{array}\right.
\end{equation} 

\begin{equation}\label{eqn:SPC-parity}
  parity = \mathlarger{\mathlarger{\mathlarger{\oplus}}}_{i=0}^{N_S-1} \text{HD}_i
\end{equation}

\noindent
If the parity constraint is satisfied, the decoding of the SPC node is assumed successful. If the parity is not satisfied, it means that there is at least one error. To satisfy the parity check constraint, the bit with the least reliable LLR is found (Eq. \ref{eqn:SPC-argmin}) and flipped (Eq. \ref{eqn:SPC-final}):

\begin{equation}\label{eqn:SPC-argmin}
  j = \arg\min(|\alpha_i|); ~0 \leq i < N_S,
\end{equation} 

\begin{equation}\label{eqn:SPC-final}
  {\beta}_i=\left\{
  \begin{array}{@{}ll@{}}
    \text{HD}_i \oplus parity, & \text{when}~ i = j \text{,}  \\
    \text{HD}_i, & \text{otherwise.}
  \end{array}\right.
\end{equation}  

Merging schemes for these special nodes were also proposed in the Fast-SSC decoder to improve the throughput further. A complete list of operations are detailed in Table \ref{tab:fast-instr}, including their merged operations.

\subsubsection{Decoder Architecture}\label{sec:fastssc_arch}
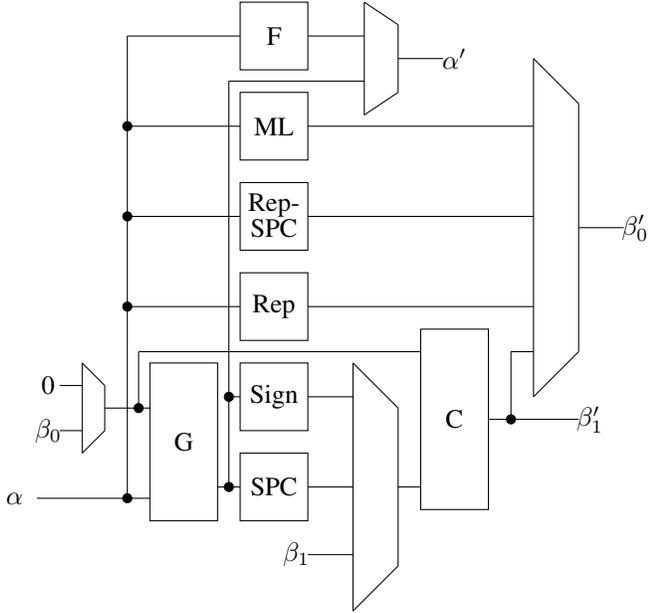
\begin{figure}
  \centering
  \scalebox{1.00}{
  \begin{tikzpicture}[scale=.60]
  \node [color=black] at (0,0) {$\alpha$} ;
  \draw [-] (0.5,0) -- (3.0,0);
  \draw [-] (2.5,0.0) -- (2.5,10.25);
 
  \draw [fill] (2.5,0) circle [radius=.1];
  \draw (3.0,-0.5) rectangle (4.5,3);
  \node [color=black] at (3.75,1.25) {G};
  \draw [fill] (2.75,2.0) circle [radius=.1];
 
  \draw [-] (2.0,2) -- (3.0,2);
  \draw [-] (2.75,2) -- (2.75,3.25);
  \draw [-] (2.75,3.25) -- (9.00,3.25);
  \draw [-] (2.0,2.5) -- (2.0,1.5);
  \draw [-] (1.5,3.0) -- (2.0,2.5);
  \draw [-] (1.5,1.0) -- (1.5,3.0);
  \draw [-] (2.0,1.5) -- (1.5,1.0);
  \draw [-] (1.0,2.5) -- (1.5,2.5);
  \draw [-] (1.0,1.5) -- (1.5,1.5);
  \node [color=black] at (0.75,2.5) {0};
  \node [color=black] at (0.75,1.5) {$\beta_0$};

  \draw (5.0,-0.5) rectangle (6.5,1.0);
  \node [color=black] at (5.75,0.25) {SPC};
  \draw (5.0,1.5) rectangle (6.5,3.0);
  \node [color=black] at (5.75,2.25) {Sign};

  \draw [-] (4.5,0.25) -- (5.0,0.25);
  \draw [fill] (4.75,0.25) circle [radius=.1];
  \draw [-] (4.75,2.25) -- (5,2.25);
  \draw [fill] (4.75,2.25) circle [radius=.1];
  \draw [-] (4.75,0.25) -- (4.75,9.25);

  \draw (5.0,3.5) rectangle (6.5,5.0);
  \node [color=black] at (5.75,4.25) {Rep};
  \draw [-] (5.0,4.25) -- (2.5,4.25);
  \draw [fill] (2.5,4.25) circle [radius=.1];
  \draw (5.0,5.5) rectangle (6.5,7.0);
  \node [color=black] at (5.75,6.50) {Rep-};
  \node [color=black] at (5.75,6.00) {SPC};
  \draw [-] (5.0,6.25) -- (2.5,6.25);
  \draw [fill] (2.5,6.25) circle [radius=.1];
  \draw (5.0,7.5) rectangle (6.5,9.0);
  \node [color=black] at (5.75,8.25) {ML};
  \draw [-] (5.0,8.25) -- (2.5,8.25);
  \draw [fill] (2.5,8.25) circle [radius=.1];
  
  \draw (5.0,9.5) rectangle (6.5,11.0);
  \node [color=black] at (5.75,10.25) {F};
  \draw [-] (5.0,10.25) -- (2.5,10.25);
  
  \draw [-] (7.5,3.0) -- (8.5,2.0);
  \draw [-] (8.5,2.0) -- (8.5,-1.5);
  \draw [-] (8.5,-1.5) -- (7.5,-2.5);
  \draw [-] (7.5,-2.5) -- (7.5,3.0);
  \draw [-] (6.5,0.25) -- (7.5,0.25);
  \draw [-] (6.5,2.25) -- (7.5,2.25);
  \draw [-] (6.5,-1.25) -- (7.5,-1.25);
  \node [color=black] at (6.25,-1.25) {$\beta_1$};
  \draw [-] (8.5,0.25) -- (9.0,0.25);
  
  \draw (9.0,-0.25) rectangle (10.5,3.75);
  \node [color=black] at (9.75,1.75) {C};
  \draw [-] (10.5,1.75) -- (12.5,1.75);
  \node [color=black] at (12.75,1.75) {$\beta_1'$};
  
  \draw [-] (4.75,9.25) -- (7.75,9.25); 
  \draw [-] (6.5,10.25) -- (7.75,10.25);  
  \draw [-] (7.75,8.5) -- (7.75,11.00);
  \draw [-] (7.75,11.00) -- (8.5,10.50);
  \draw [-] (8.5,10.50) -- (8.5,9.00);
  \draw [-] (8.5,9.0) -- (7.75,8.5);
  
  \draw [-] (8.5,9.75) -- (9.5,9.75); 
  \node [color=black] at (9.75,9.75) {$\alpha'$};
  
  \draw [-] (6.5,4.25) -- (11.5,4.25);
  \draw [-] (6.5,6.25) -- (11.5,6.25);
  \draw [-] (6.5,8.25) -- (11.5,8.25);
  \draw [-] (11.00,1.75) -- (11.00,3.25);
  \draw [fill] (11.00,1.75) circle [radius=.1];
  \draw [-] (11.00,3.25) -- (11.50,3.25);
  \draw [-] (11.5,9.75) -- (11.5,2.25);
  \draw [-] (11.5,2.25) -- (12.5,3.25);
  \draw [-] (12.5,3.25) -- (12.5,8.75);
  \draw [-] (12.5,8.75) -- (11.5,9.75);
  \draw [-] (12.5,6.00) -- (13.5,6.00);
  \node [color=black] at (13.75,6.00) {$\beta_0'$};
  
\end{tikzpicture}}
  \caption{Fast-SSC datapath architecture from \cite{sarkis14}.}
  \label{fig:fastssc}
\end{figure}

The Fast-SSC architecture described in \cite{sarkis14} contains separate memory units for channel LLR values, intrinsic 
LLR values $\alpha$, partial sums $\beta$, decoding instructions, and the final codeword. Words from channel, $\alpha$ and 
$\beta$ memory units are routed to an ALU unit, where the identified operations listed in Table \ref{tab:fast-instr} 
are performed. Left operation (F), right operation (G) and combine operation (C) are adopted from the SC decoding of 
\cite{arikan09}. The notations 0, 1 and R represent child nodes with Rate-0, Rate-1 and Rate-R (0 $<$ R $<$ 1). The 
operations with P- notation represent the operations performed without explicitly visiting the right child node. Routing 
of the memory and configuration of the datapath are controlled based on the instruction list that is compiled offline. 
The datapath for the Fast-SSC decoder is depicted in Fig. \ref{fig:fastssc}.

\begin{figure}
  \centering
  \scalebox{1.00}{
  \includegraphics[scale=0.175]{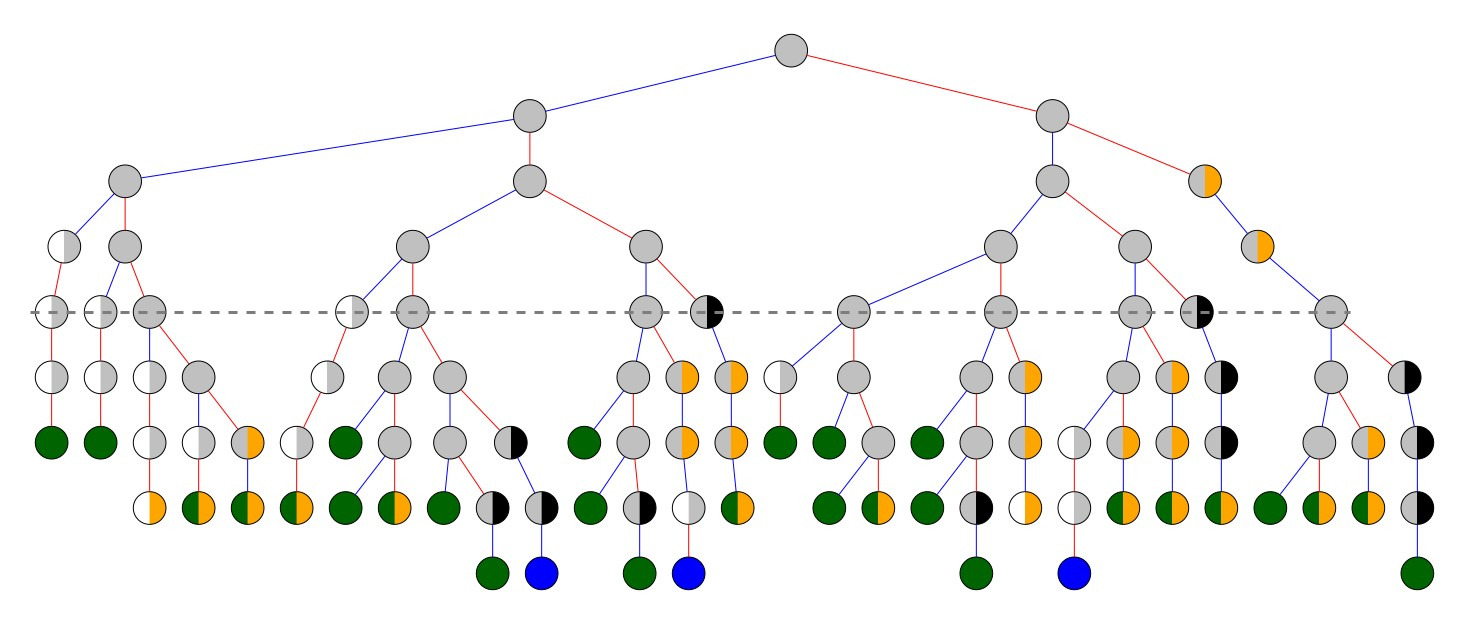}}
  \caption{$PC(1024,512)$ tree \cite{38.212} with special nodes from Table \ref{tab:fast-instr}. Green circles are 
  Rep nodes, yellow circles are SPCs, blue circles are ML nodes, white are Rate-0, black are Rate-1, gray are Rate-R. Dashed line represents the parallelization threshold when $P_e =2^6$, under which the computational resources are under-utilized.}
  \label{fig:1ktree}
\end{figure}

\begin{table}
\centering
\caption{List of operations for the Fast-SSC architecture \cite{sarkis14}.}
\label{tab:fast-instr}
\setlength{\extrarowheight}{1.1pt}
\resizebox{\columnwidth}{!}{%
\begin{tabular}{@{}rl@{}}
\toprule
Name & Operation \\
\cmidrule(r){1-1}\cmidrule(l){2-2}
F       & Operation in (\ref{eqn:alphaleft}). \\
G       & Operation in (\ref{eqn:alpharight}). \\
G0      & Operation in (\ref{eqn:alpharight}), but with $\boldsymbol{\beta^l} = \mathbf{0}$. \\
C       & Combine $\boldsymbol{\beta^l}$ and $\boldsymbol{\beta^r}$. \\
C0      & Combine $\boldsymbol{\beta^l}$ and $\boldsymbol{\beta^r}$, with $\boldsymbol{\beta^l} = \mathbf{0}$. \\
P-R1    & Calculate hard decision with $\boldsymbol{\beta^r} = \mathbf{1}$. \\
P-01    & Calculate hard decision with $\boldsymbol{\beta^l} = \mathbf{0}$ and $\boldsymbol{\beta^r} = \mathbf{1}$. \\
P-RSPC  & Calculate hard decision with right child being and SPC node. \\
P-0SPC  & Same as P-RSPC, but with $\boldsymbol{\beta^l} = \mathbf{0}$. \\
ML      & Exhaustive-search ML decoding. \\
Rep     & Calculate hard decision with (\ref{eqn:rep}). \\
RepSPC & Calculate hard decision merging Rep and SPC nodes. \\
\bottomrule
\end{tabular}}
\end{table}

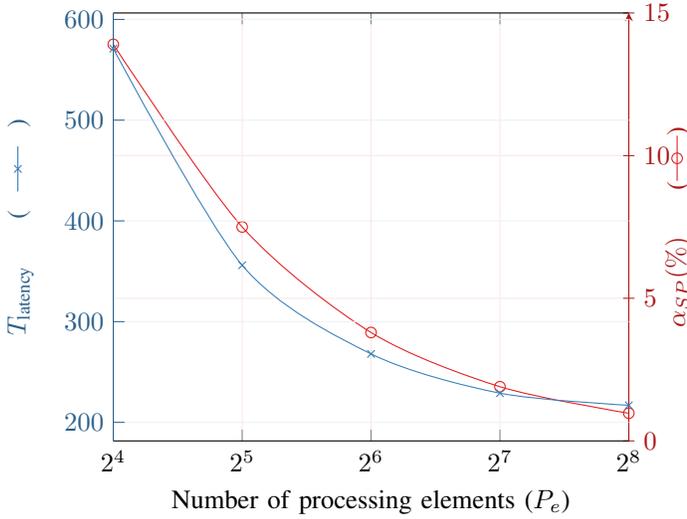
\begin{figure}[h]
\centering
\begin{tikzpicture}
     \pgfplotsset{
        y axis style/.style={
           yticklabel style=#1,
           ylabel style=#1,
           y axis line style=#1,
           ytick style=#1
        }
    }

    \begin{axis}[   axis y line=right,
                    y axis style=Set1-1!75!black,
                    ylabel style={yshift=0.5cm},
                    ymin=0, ymax=15,
                    xlabel= Number of processing elements ($P_e$),
                    xtick={16,32,64,128, 256},
                    xmin=16, xmax = 256,
                    ylabel= $\alpha_{SP} (\%)$  ~~~~(\ref{s}),
                    grid=both, grid style={Set1-1!10},
                    xmode=log,
              log basis x={2}
                ]

        \addplot[smooth,mark=o,Set1-1] coordinates  {(16 ,13.9)
                                                     (32 ,7.5)
                                                     (64 ,3.8)
                                                     (128,1.9 )
                                                     (256, 0.97)
                                                    }; \label{s}
    \end{axis}
     \begin{axis}[  axis y line*=left,
                    axis x line=none,
                    ylabel= $T_\text{latency}$ ~~~~(~~\ref{f}~~) ,
                    y axis style=Set1-2!75!black,
                    grid=both, grid style={Set1-2!10},
                    xmin=16, xmax = 256,
                    xmode=log,
              log basis x={2}
                ]

        \addplot[smooth,mark=x,Set1-2] coordinates  {(16 ,571  )
                                                     (32 ,356 )
                                                     (64 ,268 )
                                                     (128,229 ) 
                                                     (256,217 )
                                                    }; \label{f}

    \end{axis}
\end{tikzpicture}
\caption{Resource utilization rate ($\theta_{SP}$) and calculated latency ($T_\text{latency}$) in clock cycles as functions of the number of processing elements under Fast-SSC decoding for $PC(1024,512)$.}
\label{fig:pe}
\end{figure}

Execution of a single operation is called a step, and each step may take one or more clock cycles, based on the tree 
stage and the number of physical processing units dedicated to perform the operation.  If $P_e$ is too small, operations to decode a codeword takes too many clock cycles which results in reduced throughput. The number of cycles to decode a single frame decreases with increasing $P_e$, which helps increase the throughput. For example, for $PC(1024,512)$, number of clock cycles to decode a codeword is $571$ when $P_e = 16$, and is $217$ when $P_e = 256$. 
On the other hand, with increasing $P_e$, the idle time of the computational resources increase, decreasing the resource utilization. Fast-SSC decoder is a special decoder that is based on semi-parallel SC decoder family \cite{sarkis14}. According to \cite{SPSC13}, the utilization rate ($\theta_{SP}$) of a semi-parallel decoder is given by 

\[\theta_{SP} =\frac{\log_2N}{4\times P_e + \log_2\frac{N}{4P_e}},\] which leads to $\theta_{SP} = 13.8$ when $P_e = 16$ and to $\theta_{SP} = 0.9$ when $P_e = 256$, respectively. Fig.~\ref{fig:pe} plots the number of cycles to decode a codeword ($T_\text{latency}$) and resource utilization ($\theta_{SP}$) as a function of the number of processing elements. It can be seen that the utilization rate decreases significantly with increasing $P_e$, whereas after $P_e = 64$, the latency improvement is marginal. 
Thus, while keeping a reasonable resource utilization and maintaining low decoding latency, $P_e = 2^6$ is a reasonable choice for N = 1024. 

The SC tree for a $PC(1024,512)$ with nodes from Table \ref{tab:fast-instr} is presented in Fig. \ref{fig:1ktree}. The polar code construction is obtained from the 5G standard \cite{38.212}. The Fast-SSC from \cite{sarkis14} executes a total number of 212 steps in 268 clock cycles to decode a single frame of Fig. \ref{fig:1ktree} when $P_e = 2^6$. 

\subsubsection{Memory} \label{sec:BG-mem}
In the SC tree there are $\log_2(N)+1$ stages, with the highest stage $\log_2(N)$ corresponding to the root node, 
and the lowest stage $\log_2(1) = 0$ to the leaf nodes. For each stage of the SC tree, both $\alpha$ and $\beta$ are stored 
in designated memory modules. The $\alpha$ memory is comprised of two banks, each of which is $P_e$ LLRs wide. The 
partial sum $\beta$ memory is composed of two memory units, each of which has two banks, and each bank is $P_e$ bits 
wide. A full word is defined as the content from an index of a memory unit. At each clock cycle, a full word can be read 
from each memory. The F and G modules can process $2 \times P_e$ $\alpha$ and $P_e \beta$ values at once to generate $P_e$ $\alpha$ 
outputs, thus a half word can be written back to the $\alpha$ memory. Combine (C) unit takes a half word from each $\beta$ memory 
bank, to produce a full word that should be written to either of the banks. Consequently, a half word can be read from 
each $\beta$ memories, and a full word can be written back to one $\beta$ memory. 

\begin{figure}
  \centering
  \scalebox{1.00}{
  \begin{tikzpicture}[scale=.67]
\usetikzlibrary{patterns}
 \draw (0.00,0.00) rectangle (12.00,5.00);
 \draw [-,dashed] (1.00,0.00) -- (1.00,5.00);
 \draw [-,dashed] (2.00,0.00) -- (2.00,5.00);
 \draw [-,dashed] (3.00,0.00) -- (3.00,5.00);
 \draw [-] (4.00,0.00) -- (4.00,5.00);
 \draw [-,dashed] (5.00,0.00) -- (5.00,5.00);
 \draw [-] (6.00,0.00) -- (6.00,5.00);
 \draw [-] (7.00,0.00) -- (7.00,5.00);
 \draw [-] (8.00,0.00) -- (8.00,5.00);
 \draw [-] (9.00,0.00) -- (9.00,5.00);
 \draw [-] (10.00,0.00) -- (10.00,5.00);
 \draw [-] (11.00,0.00) -- (11.00,5.00);

\node [color=black, rotate=90] at (0.50,-1.00) {$S=9$};
\node [color=black, rotate=90] at (4.50,-1.00) {$S=8$};
\node [color=black, rotate=90] at (6.50,-1.00) {$S=7$};
\node [color=black, rotate=90] at (7.50,-1.00) {$S=6$};
\node [color=black, rotate=90] at (8.50,-1.00) {$S=5$};
\node [color=black, rotate=90] at (9.50,-1.00) {$S=4$};
\node [color=black, rotate=90] at (10.50,-1.00) {$S=3$};
\node [color=black, rotate=90] at (11.50,-1.00) {$S=2$};

\draw [<->] (-0.50,0.00) -- (-0.50,5.00);
\node [color=black, rotate=90] at (-1.00,2.50) {$2 \times P_e$};
 
\draw [<->] (0.00,5.50) -- (4.00,5.50);
\node [color=black] at (2.00,6.00) {$N_v / 2P_e$};

\draw [pattern=north west lines, pattern color=gray] (7.00,2.50) rectangle (8.00,5.00);
\draw [pattern=north west lines, pattern color=gray] (8.00,1.25) rectangle (9.00,5.00);
\draw [pattern=north west lines, pattern color=gray] (9.00,0.625) rectangle (10.00,5.00);
\draw [pattern=north west lines, pattern color=gray] (10.00,0.30) rectangle (11.00,5.00);
\draw [pattern=north west lines, pattern color=gray] (11.00,0.15) rectangle (12.00,5.00);

\end{tikzpicture}}
  \caption{Memory architecture (for both $\alpha$ and $\beta$) for a polar code of length $N=1024$, with $P_e = 64$. 
  $N_v$ is the node size at stage $S$ length, shaded area is unused.}
  \label{fig:memarch-old}
\end{figure}
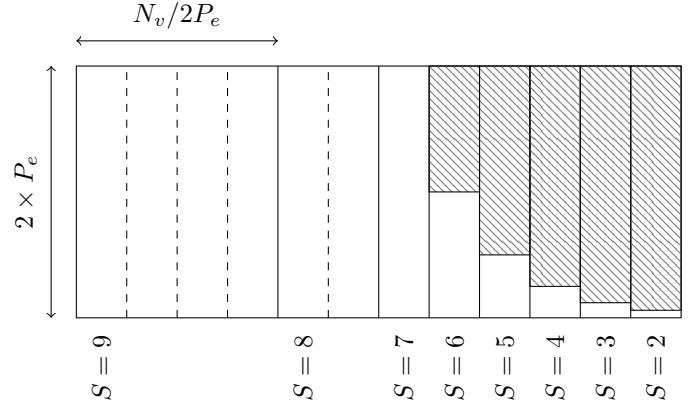

In terms of both $\alpha$ and $\beta$ memories, at least one word is reserved for each stage of the tree. Fig. 
\ref{fig:memarch-old} represents the memory architecture for both $\alpha$ and $\beta$ memories for $PC(1024,K)$ code. 
The root node (S=10) is stored explicitly in the channel memory, because it has a different quantization scheme than the 
internal LLRs. Due to special node decoding techniques mentioned in Section \ref{sec:fastssc-old}, the lower limit stage is S=2, 
thus no additional memory is required after stage S=2. In general, given that each word holds $2 \times P_e$ elements for $\alpha$ and $\beta$ memories, the number of words for each memory module is:
\begin{equation}
 \sum_{S=2}^{\log_2(N-1)} \left\lceil \frac{2^{S-1}}{P_e}\right\rceil. 
\end{equation}

\section{Improving Memory Utilization}\label{sec:proposed-memory}

As it was previously mentioned in Section~\ref{sec:fastssc-old}, one or more words are reserved in both $\alpha$ and $\beta$ memories per decoding stage in the tree. The word size is decided by the number of processing elements $P_e$, that acts as a parallelization factor. $P_e$ can also be interpreted as a threshold on the SC tree (dashed line in Fig. \ref{fig:1ktree}), where the node size $N_v=P_e$. The stages above this parallelization threshold are collectively called high-stage, and each stage in high-stage fully utilizes the dedicated memory words for that level; below the threshold (low-stage), only a portion of the memory is used. In fact, the total number of variables used at low-stage can be expressed as
\begin{equation}
\sum_{\text{S}=2}^{\log_2 P_e} 2^\text{S} < 2 \times P_e
\end{equation}
which can fit into a single memory word.

\begin{figure}
  \centering
  \scalebox{1.00}{
  \begin{tikzpicture}[scale=.70]
\usetikzlibrary{patterns}

\draw (0.00,0.00) rectangle (12.00,1.00);

\draw [-] (2.00,0.00) -- (2.00,1.00);
\draw [-] (4.00,0.00) -- (4.00,1.00);
\draw [-] (6.00,0.00) -- (6.00,1.00);
\draw [-] (8.00,0.00) -- (8.00,1.00);
\draw [-] (10.0,0.00) -- (10.00,1.00);

\node [color=black] at (1.00,-0.40) {$S=6$};
\node [color=black] at (3.00,-0.40) {$S=5$};
\node [color=black] at (5.00,-0.40) {$S=4$};
\node [color=black] at (7.00,-0.40) {$S=3$};
\node [color=black] at (9.00,-0.40) {$S=2$};

\draw [<->] (0.00,1.20) -- (2.00,1.20);
\draw [<->] (2.00,1.20) -- (4.00,1.20);
\draw [<->] (4.00,1.20) -- (6.00,1.20);
\draw [<->] (6.00,1.20) -- (8.00,1.20);
\draw [<->] (8.00,1.20) -- (10.00,1.20);
\draw [<->] (10.00,1.20) -- (12.00,1.20);

\node [color=black] at (1.00,1.50) {$P_e$};
\node [color=black] at (3.00,1.50) {$P_e/2$};
\node [color=black] at (5.00,1.50) {$P_e/4$};
\node [color=black] at (7.00,1.50) {$P_e/8$};
\node [color=black] at (9.00,1.50) {$P_e/16$};
\node [color=black] at (11.00,1.50) {$P_e/16$};

\draw [pattern=north west lines, pattern color=gray] (10.00,0.00) rectangle (12.00,1.00);

\end{tikzpicture}}
  \caption{Proposed memory structure below the parallelization threshold, for $PC(1024,K)$ and $P_e = 64$.}
  \label{fig:memarch-new}
\end{figure}
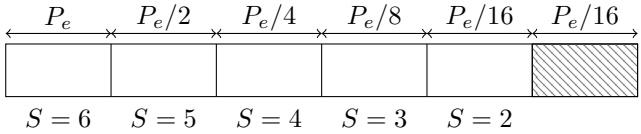

We improve the memory utilization by storing all the variables relative to the low-stage into a single memory word in both $\alpha$ and $\beta$ memory units. Fig.~\ref{fig:memarch-new} describes the new memory configuration for the low-stage, using a polar code $PC(1024,K)$ and $P_e=64$ as an example. With the proposed solution, the number of $2\times Pe$-sized words for each memory becomes
\begin{equation}
\sum_{S=\log_2(P_e)}^{\log_2(N-1)} \left\lceil \frac{2^{S-1}}{P_e}\right\rceil
\end{equation}
Considering a polar code $PC(1024,K)$ and $P_e=64$, the memory utilization increases from $66\%$ to $99.6\%$.

With these modifications to the memory structure, when a node at the low-stage is to be processed, the entire last word is fetched from the memory. Since the content of the memory is relative to multiple stages, executing multiple operations per clock cycle becomes possible.

\section{Operation Merging}\label{sec:proposed-merging}

In this work, we define an operation as a \textit{leaf operation} if it involves a leaf node estimation, and as a \textit{branch operation} if it does not include any bit estimations. According to this classification, hard decision (\ref{eqn:beta-leaf}), Rate-0, Rate-1, Rep (\ref{eqn:rep}) and SPC (\ref{eqn:SPC-parity})-(\ref{eqn:SPC-final}) calculations are leaf node operations, whereas 
F (\ref{eqn:alphaleft}), G (\ref{eqn:alpharight}) and C (\ref{eqn:beta}) are branch operations for SC decoding. Note that proposed merged operations do not affect the error-correction performance.

\subsection{Merging Branch Operations}\label{subsec:proposed-branchmerging}

\begin{figure}
 \centering
 \includegraphics[width=\columnwidth]{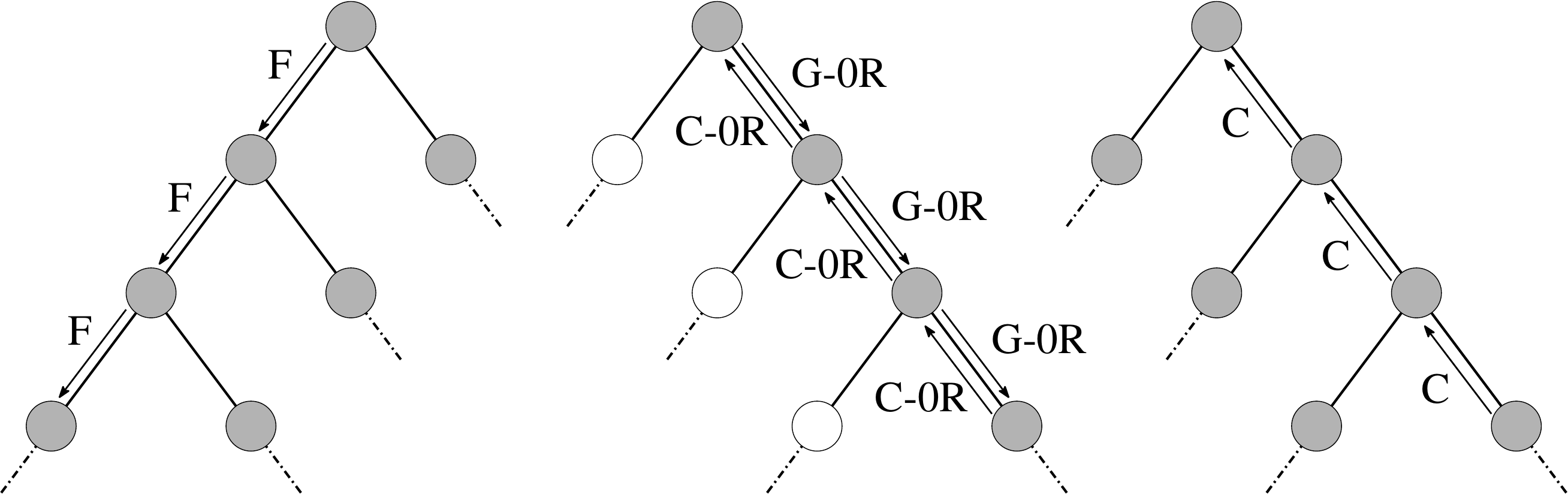}
 \caption{Required conditions to perform multiple F, G0, C and C0 operations on a polar code decoder tree.}
 \label{fig:merged_f_g_c}
\end{figure}

If the memory configuration in Fig.~\ref{fig:memarch-new} is used within the Fast-SSC decoder architecture, all $\alpha$ and $\beta$ variables below the parallelization threshold becomes available for processing at the same time. This enables the Fast-SSC processor to perform multiple operations at a single cycle. In other words, operations below the parallelization threshold are available for merging. However, the impact of operation merging on system critical path should be minimized and thus the original critical path should be considered as an upper delay bound while performing multiple low-stage operations in a single cycle. It was observed that the critical path of the original Fast-SSC architecture is determined by the SPC node. Compared to SPC-related operations, branch operations introduce a significantly lower delay; this provides the opportunity to merge them without increasing the system critical path. Consequently, we exploit the branch operation merging opportunities at low-stage. Based on the data dependencies while decoding, the following merging scenarios are possible for operations of the same kind: 

\begin{itemize}
\item Multiple F operations: The traversal of the SC tree has a left branch priority, which enables to perform multiple F operations consecutively.
\item Multiple G0 operations: Tree traversal allows consecutive G operations only when the left node is a Rate-0 node and needs not to be traversed.
\item Multiple C/C0 operations: A sequence of Combine operations is possible when the operation ascends from a right branch, i.e. $\beta$ values of the left children are already available. This constraint does not apply to the last C operation in the sequence.
\end{itemize}
These four different merging branch operations of the same kind are visualized in Fig.~\ref{fig:merged_f_g_c}.

The combination of different branch operations at low-stage is also feasible. It was observed that a G operation is often followed by an F operation, which can be merged together to form a new operation called G-F. Similar observations were made for F-G0, C-G and C0-G. A complete list of merged branch operations and their associated potential step reduction is presented in Table~\ref{tab:potentials-branch} for $PC(1024,512)$ \cite{38.212}. According to Table~\ref{tab:potentials-branch}, the amount of time step reduction increases with $P_e$. It can also be observed that G-F merging scenario returns the most amount of reduction. Note that, the merging scenarios in Table~\ref{tab:potentials-branch} are computed independently, without considering any conflicts between the merging scenarios. A set of guidelines for how to merge operations are detailed in Section~\ref{subsec:guidelines}.

\begin{table}[]
\centering
\caption{Potential time step reduction with merging branch operations of the same or different kinds, for $PC(1024,512)$ \cite{38.212}.}
\label{tab:potentials-branch}
\resizebox{\columnwidth}{!}{%
\begin{tabular}{l rrr }
\toprule
Merging Scenario & $P_e = 32$ & $P_e = 64$ & $P_e = 128$ \\
\cmidrule(r){1-1} \cmidrule(l){2-2}\cmidrule(l){3-3} \cmidrule(l){4-4}
F$^{\times2}$    & $2.78\%$   & $6.34\%$   & $9.09\%$  \\
F$^{\times3}$    & $0.56\%$   & $2.99\%$   & $6.06\%$  \\
F$^{\times4}$    & ---        & ---        & $1.30\%$  \\
G0$^{\times2}$   & $0.83\%$   & $2.24\%$   & $3.03\%$  \\
G0$^{\times3}$   & ---        & $0.75\%$   & $1.73\%$  \\
C$^{\times2}$    & $1.94\%$   & $3.73\%$   & $5.63\%$  \\
C$^{\times3}$    & $0.56\%$   & $2.24\%$   & $3.46\%$  \\
C$^{\times4}$    & ---        & ---        & $1.30\%$  \\
C0$^{\times2}$   & $0.67\%$   & $2.61\%$   & $3.03\%$  \\
C0$^{\times3}$   & $0.56\%$   & $1.49\%$   & $1.73\%$  \\
\midrule
G-F              & $5.00\%$   & $8.58\%$   & $11.26\%$ \\
F-G0             & $0.83\%$   & $1.87\%$   & $3.03\%$  \\
C-G              & $1.94\%$   & $3.73\%$   & $5.19\%$  \\
C0-G             & $1.11\%$   & $2.24\%$   & $3.03\%$  \\
\bottomrule
\end{tabular}}
\end{table}

\subsection{Merging Special Nodes}\label{subsec:proposed-leafmerging}

\begin{table*}[]
\centering
\caption{Potential time step reduction with merging different special nodes from \cite{sarkis14}, for $PC(1024,512)$ \cite{38.212} and $P_e \geq 16$.}
\label{tab:potentials-leaf}
\resizebox{1.6\columnwidth}{!}{%
\hspace{-5em}\begin{tabular}{l l l rrr }
\toprule
Merging Scenario & New Node Size & Translation & $P_e = 32$ & $P_e = 64$ & $P_e = 128$ \\
\cmidrule(r){1-1} \cmidrule(l){2-2}\cmidrule(l){3-3} \cmidrule(l){4-4}\cmidrule(l){5-5}\cmidrule(l){6-6}
Rep-RepSPC    & $N_v = 16$ & F $\rightarrow$ Rep $\rightarrow$ G $\rightarrow$ RepSPC $\rightarrow$ C & $3.33\%$ & $4.48\%$ & $5.19\%$ \\
Rate0-RepSPC  & $N_v = 16$ & G0 $\rightarrow$ RepSPC $\rightarrow$ C0 & $1.11\%$ & $1.49\%$ & $1.73\%$ \\
RepSPC-Rate1 & $N_v = 16$ & F $\rightarrow$ RepSPC $\rightarrow$ P-R1 &  $0.28\%$ & $0.37\%$ & $0.43\%$ \\
Rep-Rate1  & $N_v = 8 $ & F $\rightarrow$ Rep $\rightarrow$ P-R1    & $1.11\%$ & $1.49\%$ & $1.73\%$ \\
Rate0-ML   & $N_v = 8$ & G0 $\rightarrow$ ML $\rightarrow$ C0 &  $1.11\%$ & $1.49\%$ & $1.73\%$ \\
ML-Rate1 & $N_v = 8$ & F $\rightarrow$ ML $\rightarrow$ P-R1 & $0.28\%$ & $0.37\%$ & $0.43\%$ \\
F-Rep   & $N_v \in \{8,16,32\}$ & F $\rightarrow$ Rep &  $2.78\%$ & $3.73\%$ & $4.33\%$ \\
\bottomrule
\end{tabular}}
\end{table*}

As mentioned in Section \ref{sec:fastssc-old}, the Fast-SSC architecture from \cite{sarkis14} uses merging of special nodes in order to improve the throughput. For example, the SPC and Sign operators in between G and C modules in Fig. \ref{fig:fastssc} enable the datapath to execute P-RSPC, P-0SPC, P-R1, P-01, G0 and C0 operations from Table \ref{tab:fast-instr}. On the other hand, a separate module for RepSPC node is instantiated to avoid the critical path. In this Section, based on the observations made from the polar code tree in Fig.~\ref{fig:polartree}, we identify all possible special node merging scenarios. It should be noted that some of the special node merging scenarios in this Section fall within the generalized nodes from \cite{ardakani-sc}, where however no hardware implementation were proposed.

Table~\ref{tab:potentials-leaf} describes a number of possible leaf node merging scenarios along with their node sizes, breakdown of operations (translations), and amount of time step reduction with respect to $P_e$. It is important that the $P_e$ threshold must be larger than $16$ to support the new node sizes in Table~\ref{tab:potentials-leaf}. If $P_e \leq 16$, described merging operations fall above the parallelization threshold in the polar code tree, and they cannot be merged. Rep-Rate1 and Rate0-RepSPC nodes were previously identified in \cite{giardJSPS}; we include them in this work in order to compare them with newly identified nodes in terms of time step reduction. Similar to Table~\ref{tab:potentials-branch}, the time step reduction calculations are done independently from each other. According to Table~\ref{tab:potentials-leaf}, merged special node operation Rep-RepSPC returns the most potential time step reduction, followed by F-Rep. Note that the instances of F-Rep operation occurs within Rep-RepSPC nodes, which can be observed in their translations.

The selection of merging scenarios for special nodes should not only be performed with respect to their independent contribution on time step reduction, but also with respect to their impact on the maximum operating frequency. Merging a special node with a Rate-0 node is more favorable than merging with Rate-1 nodes since their calculation takes less time and area. The impact of merged operations on maximum operating frequency should be taken into account when compiling a new instruction set.

\subsection{Guidelines for Operation Merging}\label{subsec:guidelines}
The time step reduction amounts in Table \ref{tab:potentials-branch} and in Table \ref{tab:potentials-leaf} are observed independently: not all of the merging can be exploited at the same time. For example, if a series of operations such as (G-F-F) is present at low-stage, the potential time step reduction consider both \{G-F;F\} and \{G-F$^{\times 2}$\}, but only one scheme can be implemented at once. In order to minimize the number of conflicts between different operation merging schemes and to maximize the time step reduction, we developed a merging algorithm based on the following observations. Note that a tail operation refers to the last operation, and a head operation refers to the first operation in an operation sequence or subsequence. 
\begin{itemize}
\item G-F vs. F$^{\times\{2,3,4\}}$: When a series of F operations are observed in the original instruction list, they are merged starting from the tail F operation, in order to minimize conflict with the possible merging of a G-F operation.
\item (C/C0)-G-F sequence: It was observed that in the polar code decoding sequence, the operation flow for C-G or C0-G operation is always followed by an F operation, if not followed by a special node. This is because the node after C-G (C0-G) sequence is always the root node of an an unexploited subtree. In addition, the G-F operation occurs more frequently than C-G/C0-G operations. Finally, the C/C0 operation can be merged within another operation such as  C$^{\times 2}$. As a result, although C-G/C0-G is a possible operation merging scenario, it is not used in our approach.
\item C-G vs. C$^{\times\{2,3,4\}}$ (C0-G vs. C0$^{\times\{2,3\}}$): If C-G/C0-G operation will be used as a merging scenario, the merging of consecutive C (C0) operations is advised to begin from the head operation towards the tail operation, to maximize the chances of merging with a G operation that follows it, leading to a C-G (C0-G) operation.
\item F-G0 vs. G0$^{\times\{2,3\}}$ and F$^{\times\{2,3,4\}}$: Merging of consecutive G0 operations starts from the tail G0 operation towards the head operation, to minimize the conflict with potential F-G0 operation. To further minimize conflicts, F operations should be merged starting from the head operation; however this decision would contradict the decision made for the sake of maximizing G-F operations. Based on observations, the number of occurrences of G-F is much higher than that of F-G0, thus F is decided to be merged starting from the tail F operation.
\item F-Rep vs. F$^{\times\{2,3,4\}}$: To minimize the conflict between F-Rep and consecutive F operations, merging F is advised to begin from the head operation. However, similar to the previous case, this conflicts with the G-F operation merging scenario. Although F-Rep is one of the leaf node operations that returns a favorable amount of time step reduction, if Rep-RepSPC node is used, most of F-Rep operations will be included within it. Because of this, F-Rep operation loses priority against G-F operation. Consequently, merging F operations begin from their tail operation, and F-Rep should be merged from the remaining ones.
\item Merged special nodes vs. F/G0/C/C0: As described in Table~\ref{tab:potentials-leaf}, merging special nodes include branch operations. Additionally, merging special nodes returns more time step reduction compared to branch operation merging. Thus, in order to minimize merging conflict and maximize savings, special nodes must be merged before merging branch operations.
\item Merging multiple branch operations: It was observed that if the leaf nodes are merged first, followed by merging branch operations of different kinds (e.g. G-F), the impact of merging operations of the same kind on time step savings reduces dramatically; on the other hand, the benefits of merging leaf nodes and branch operations of different kinds are maximized. Hence, given an instruction sequence derived from Table~\ref{tab:fast-instr}, leaf nodes are merged first, followed by merging branch operations of different kinds, and finally merging branch operations of the same kind.
\end{itemize}

\begin{table}
\centering
\caption{New instruction set for the proposed Fast-SSC decoder.}
\label{tab:instr:new}
\setlength{\extrarowheight}{1.1pt}
\resizebox{\columnwidth}{!}{%
\begin{tabular}{l  l}
\toprule
Operation & Details \\
\cmidrule(r){1-1} \cmidrule(l){2-2}
F$^{\times 2}$ & Two consecutive F operations. \\
G0$^{\times 2}$ & Two consecutive G0 operations. \\
C$^{\times 2}$, C$^{\times 3}$ & Up to three consecutive C operations. \\
C0$^{\times 2}$, C0$^{\times 3}$ & Up to three consecutive C0 operations. \\
G-F & G operation followed by an F operation. \\
F-G0 & F operation followed by a G0 operation. \\
F-Rep & F operation followed by a Rep node.\\
Rep-RepSPC & Node with Rep and RepSPC nodes as its children.\\
Rep-Rate1 & Node with Rep and Rate-1 nodes as its children.\\
Rate0-ML & Node with Rate-0 and ML nodes as its children.\\
\bottomrule
\end{tabular}}
\end{table}

Based on the observations above, a new instruction set is derived for our proposed Fast-SSC decoder. The proposed algorithm uses the original Fast-SSC instructions of Table \ref{tab:fast-instr} along with the newly identified instructions in Table~\ref{tab:instr:new}. In order to maintain a reasonable operating frequency, the delay of the critical path should be maintained as much as possible. Our studies show that up to three C/C0 operations, and up to two F/G0 operations can be merged with minimum effect on the critical path delay. Note that some of the identified merged operations from Table~\ref{tab:potentials-branch} and Table~\ref{tab:potentials-leaf} are not used in Table~\ref{tab:instr:new} since either they did not occur after following the described merging guidelines, or their impact on reducing the number of time steps is negligible.

The amount of savings in terms of the number of operations and the number of time steps with the new operation set listed in Table~\ref{tab:instr:new} is detailed in Table~\ref{tab:savings} for polar codes of length $N=1024$, $R \in \{\frac{1}{4},\frac{1}{2},\frac{3}{4}\}$ and $P_e \in \{32,64,128\}$.  Note that the 5G standard allows a wide range of code rates with binary granularity. Thus, we limit our exploration within three selected rates. It can be seen that the amount of reduction in terms of both number of operations and number of time steps increases with $P_e$. It can also be observed that with increasing $P_e$, the amount of time step reduction increases at lower rate codes, since the new instruction list in Table~\ref{tab:instr:new} favors Rate-0/Rep nodes more than Rate-1/SPC nodes.

\begin{table}[]
\centering
\caption{Amount of savings in terms of number of operations and time steps, with operation set from Table~\ref{tab:instr:new} for polar codes of various rates and $P_e$ compared to Fast-SSC decoder of \cite{sarkis14}.}
\label{tab:savings}
\begin{tabular}{l l rrr}
\toprule
Polar Code & $R$ & $P_e$ & Oper. Savings & Time Step Savings \\
\cmidrule(r){1-2} \cmidrule(l){3-3} \cmidrule(l){4-4}\cmidrule(l){5-5}

\multirow{3}{*}{$PC(1024,256)$} & \multirow{3}{*}{\large $\frac{1}{4}$} & $32$  & $27.43\%$ & $15.14\%$ \\
                                &                                & $64$  & $34.86\%$ & $26.87\%$ \\
                                &                                & $128$ & $38.86\%$ & $35.42\%$ \\
\cmidrule(r){1-2} \cmidrule(l){3-3} \cmidrule(l){4-4}\cmidrule(l){5-5}
\multirow{3}{*}{$PC(1024,512)$} & \multirow{3}{*}{\large $\frac{1}{2}$} & $32$  & $26.54\%$ & $15.56\%$ \\
                                &                                & $64$  & $32.70\%$ & $25.75\%$ \\
                                &                                & $128$ & $35.55\%$ & $32.47\%$ \\
\cmidrule(r){1-2} \cmidrule(l){3-3} \cmidrule(l){4-4}\cmidrule(l){5-5}
\multirow{3}{*}{$PC(1024,768)$} & \multirow{3}{*}{\large $\frac{3}{4}$} & $32$  & $22.54\%$ & $12.42\%$ \\
                                &                                & $64$  & $26.01\%$ & $19.91\%$ \\
                                &                                & $128$ & $30.06\%$ & $27.23\%$ \\
\bottomrule
\end{tabular}
\end{table}

\section{Decoder Architecture}\label{sec:architecture}

\subsection{Architecture Overview}

\begin{figure}
  \centering
  \scalebox{1.00}{
  \begin{tikzpicture}[scale=.50]

\draw (0.00,0.00) rectangle (6.00,2.00);
\node [color=black] at (3.00,1.00) {Codeword RAM};

\draw (0.00,2.50) rectangle (6.00,4.50);
\node [color=black] at (3.00,3.50) {$\beta$-RAM};

\draw (0.00,5.00) rectangle (6.00,7.00);
\node [color=black] at (3.00,6.00) {Channel~/~$\alpha$-RAM};

\draw [dashed] (-0.30,-0.30) rectangle (6.30,7.30);

\draw (8.00,0.00) rectangle (12.00,7.00);
\node [color=black] at (10.00,3.50) {Datapath};

\draw [decoration={markings,mark=at position 1 with {\arrow[scale=1.5,>=stealth]{>}}},postaction={decorate}] (6.00,5.7) -- (8.00,5.7);
\draw [decoration={markings,mark=at position 1 with {\arrow[scale=1.5,>=stealth]{>}}},postaction={decorate}] (8.00,6.3) -- (6.00,6.3);

\draw [decoration={markings,mark=at position 1 with {\arrow[scale=1.5,>=stealth]{>}}},postaction={decorate}] (6.00,3.2) -- (8.00,3.2);
\draw [decoration={markings,mark=at position 1 with {\arrow[scale=1.5,>=stealth]{>}}},postaction={decorate}] (8.00,3.8) -- (6.00,3.8);

\draw [decoration={markings,mark=at position 1 with {\arrow[scale=1.0,>=stealth]{>}}},postaction={decorate}] (3.00,2.5) -- (3.00,2.0);

\draw (0.00,7.70) rectangle (4.00,9.70);
\node [color=black] at (2.00,8.70) {Instr. RAM};

\draw (5.00,7.70) rectangle (12.00,9.70);
\node [color=black] at (8.50,8.70) {Controller};

\draw [decoration={markings,mark=at position 1 with {\arrow[scale=1.2,>=stealth]{>}}},postaction={decorate}] (4.00,8.7) -- (5.00,8.7);

\draw [decoration={markings,mark=at position 1 with {\arrow[scale=1.0,>=stealth]{>}}},postaction={decorate}] (5.50,7.7) -- (5.50,7.3);

\draw [decoration={markings,mark=at position 1 with {\arrow[scale=1.25,>=stealth]{>}}},postaction={decorate}] (10.00,7.7) -- (10.00,7.0);

\node [color=black] at (6.00,10.20) {\footnotesize Instruction Sequence};
\draw [-](3.5,10.20)  -- (2.00,10.20);
\draw [decoration={markings,mark=at position 1 with {\arrow[scale=1.2,>=stealth]{>}}},postaction={decorate}] (2.00,10.20) -- (2.00,9.7);

\node [color=black,rotate=90] at (-0.80,3.4) {\footnotesize Channel LLR};
\draw [-](-0.80,5.2)  -- (-0.80,6.0);
\draw [decoration={markings,mark=at position 1 with {\arrow[scale=1.2,>=stealth]{>}}},postaction={decorate}] (-0.80,6.0) -- (-.00,6.0);

\end{tikzpicture}}
  \caption{High-level architecture of the proposed Fast-SSC decoder.}
  \label{fig:arch_new_highlevel}
\end{figure}
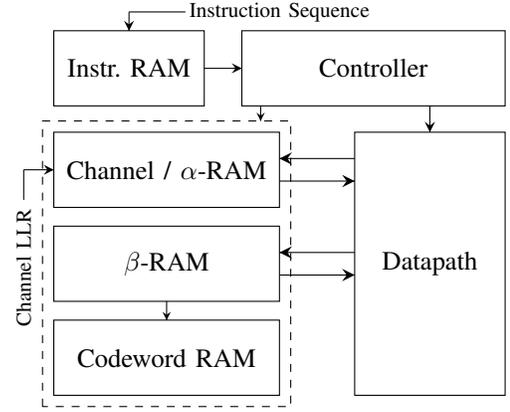

In order to evaluate the impact of merged operations on the throughput, a new Fast-SSC architecture has been implemented. The architecture supports the fast node decoding techniques from Table~\ref{tab:fast-instr} as well as the new instruction set from Table~\ref{tab:instr:new}. The high-level description of the architecture is depicted in Fig.~\ref{fig:arch_new_highlevel}. Decoding sequence begins after loading the instructions into the instruction RAM, and when Channel LLRs are present in the Channel RAM. Note that loading the instruction sequence has to be done only once. The LLR values obtained from the channel are stored in the channel memory, which has a different quantization scheme than the internal LLR $\alpha$ memory. The controller tracks the stage size for each instruction and routes the correct words to and from the $\alpha$ and $\beta$ memory units. A codeword RAM is separately instantiated from the $\beta$-RAM and stores the estimated codeword. From $\alpha$-RAM, $2\times P_e$ LLR values are fetched and $P_e$ LLRs are stored at a time. In case of $\beta$-RAM, $2\times P_e$ partial sum values can be read and stored in a single cycle. For high-stage LLR and partial sum computations, the information is processed $2 \times P_e$ elements at a time. Hence, for $S > \log_2P_e$, $2^S/2 \times P_e$ time steps are required. For low-stage operations, a single time step is required.

In merged branch operations, the output of a prior sub-operation is immediately used by the operation that follows it, and storage can often be avoided. For example, the output of the first sub-operation of G0$^{\times 2}$ is used only by its following G0 operation and is never used again. Consequently, only the output of last sub-operation is stored into memory. On the other hand, the output of each F sub-operation of F$^{\times 2}$ will be used by another operation in the future, which makes it mandatory to store the output of the first F sub-operation. By avoiding the storage of intermediate values that will not be used in future instructions, it is possible to save memory bandwidth and increase the number of parallel operations that can be performed by merged operations. The complete list of the intermediate data storing choices, and maximum parallel operations for merged operations are listed in Table \ref{tab:newinstr-details}. 

\begin{table}
\centering
\caption{Details of data storing requirements and maximum number of inputs for merged operations.}
\label{tab:newinstr-details}
\setlength{\extrarowheight}{1.1pt}
\resizebox{\columnwidth}{!}{%
\begin{tabular}{rll}
\toprule
Branch      & Storage of  & Max. number \\ 
Operations   & all data  & of inputs   \\ 
\cmidrule(r){1-1}\cmidrule(l){2-3}
F$^{\times 2}$     & Yes & $P_e$  \\
G0$^{\times 2}$    & No  & $2 \times P_e$  \\
C/C0$^{\times \{2,3\}}$  & No  & $P_e/2$  \\
G-F     & Yes & $P_e$  \\
F-G0    & No  & $2 \times P_e$  \\
C-G/C0-G & Yes & $P_e/2$  \\
\bottomrule
\end{tabular}}

\end{table}

\begin{figure}
  \centering
  \begin{tikzpicture}[scale=.53]

\draw (0.00,0.00) rectangle (2.00,4.00);
\node [color=black] at (1.00,2.00) {\Large G};

\draw (3.00,0.00) rectangle (5.00,1.50);
\node [color=black] at (4.00,0.75) {SPC};
\draw (3.00,2.50) rectangle (5.00,4.00);
\node [color=black] at (4.00,3.25) {Sign};

\draw [-](2.0,3.25)  -- (3.0,3.25);
\draw [-](2.5,3.25)  -- (2.5,0.75);
\draw [-](2.5,0.75)  -- (3.0,0.75);
\draw [fill] (2.5,3.25) circle [radius=.1];

\draw [-](5.0,3.25)  -- (5.5,3.25);
\draw [-](5.0,0.75)  -- (5.5,0.75);

\draw [-](5.5,-1.50)  -- (5.5,4.00);
\draw [-](5.5,4.00)  -- (6.5,3.50);
\draw [-](6.5,3.50)  -- (6.5,-1.00);
\draw [-](6.5,-1.00)  -- (5.5,-1.50);

\draw [-](4.5,-0.75)  -- (5.5,-0.75);
\node [color=black] at (4.0,-0.75) {$\boldsymbol{\beta_1}$};

\draw [-](6.5,2.00)  -- (7.5,2.00);
\draw [-](-1.0,4.50)  -- (7.5,4.50);
\draw [fill] (1.0,4.5) circle [radius=.1];
\draw [-](1.0,4.50)  -- (1.0,4.00);
\draw (7.50,1.50) rectangle (9.50,5.50);
\node [color=black] at (8.50,3.50) {\large C/C0};
\draw [-](9.5,2.00)  -- (10.5,2.00);
\node [color=black] at (11.00,2.00) {$\boldsymbol{\beta_1}'$};
\draw [-](9.5,4.50)  -- (10.0,4.50);

\draw (0.00,5.00) rectangle (7.0,8.00);
\node [color=black] at (2.00,7.25) {Rep-RepSPC};
\node [color=black] at (2.00,6.50) {Rep-Rate1};
\node [color=black] at (2.00,5.75) {Rate0-ML};
\draw [-](4.0,5.5)  -- (4.00,7.5);
\node [color=black] at (5.50,7.25) {RepSPC};
\node [color=black] at (5.50,6.50) {Rep};
\node [color=black] at (5.50,5.75) {ML};
\draw [-](7.0,6.50)  -- (10.0,6.5);

\draw (0.00,9.50) rectangle (2.00,11.00);
\node [color=black] at (1.00,10.25) {\large G-F};
\draw [-](2.00,10.25)  -- (7.50,10.25);

\draw (0.00,11.50) rectangle (2.00,13.00);
\node [color=black] at (1.00,12.25) {\large G0};
\draw [-](2.00,12.25)  -- (7.50,12.25);

\draw (3.00,10.50) rectangle (5.00,12.00);
\node [color=black] at (4.00,11.25) {\large F-G0};
\draw [-](5.00,11.25)  -- (7.50,11.25);

\draw (3.00,12.50) rectangle (5.00,14.00);
\node [color=black] at (4.00,13.25) {\large F};
\draw [-](5.00,13.25)  -- (7.50,13.25);

\draw (3.00,8.25) rectangle (6.00,10.00);
\node [color=black] at (4.50,9.125) {\large F-Rep};
\draw [-](6.0,8.75)  -- (10.0,8.75);

\draw [-](10.00,4.00)  -- (10.00,9.50);
\draw [-](10.00,9.50)  -- (11.00,9.00);
\draw [-](11.00,9.00)  -- (11.00,4.50);
\draw [-](11.00,4.50)  -- (10.00,4.00);
\draw [-](11.00,6.75)  -- (12.00,6.75);
\node [color=black] at (12.50,6.75) {$\boldsymbol{\beta_0}'$};

\draw [-](7.50,9.00)  -- (7.50,13.75);
\draw [-](7.50,13.75)  -- (8.50,13.25);
\draw [-](8.50,13.25)  -- (8.50,9.50);
\draw [-](8.50,9.50)  -- (7.50,9.00);
\draw [-](8.5,11.375)  -- (9.5,11.375);
\node [color=black] at (10,11.375) {$\boldsymbol{\alpha}'$};

\draw [-](-1.5,13.25)  -- (-0.5,13.25);
\node [color=black] at (-2.00,13.25) {$\boldsymbol{\alpha}$};

\draw [-](-0.5,2.00)  -- (-0.5,13.25);

\draw [fill] (-0.5,13.25) circle [radius=.1];
\draw [-](-0.5,13.25)  -- (3.0,13.25);
\draw [fill] (-0.5,11.25) circle [radius=.1];
\draw [-](-0.5,11.25)  -- (3.0,11.25);
\draw [fill] (-0.5,12.25) circle [radius=.1];
\draw [-](-0.5,12.25)  -- (0.0,12.25);
\draw [fill] (-0.5,10.50) circle [radius=.1];
\draw [-](-0.5,10.50)  -- (0.0,10.50);
\draw [fill] (-0.5,9.25) circle [radius=.1];
\draw [-](-0.5,9.25)  -- (3.0,9.25);
\draw [fill] (-0.5,6.5) circle [radius=.1];
\draw [-](-0.5,6.5)  -- (0.0,6.50);
\draw [-](-0.5,2.00)  -- (0.0,2.00);

\draw [-](-1,5.50)  -- (-1,3.50);
\draw [-](-1,3.50)  -- (-1.5,3.25);
\draw [-](-1.5,3.25)  -- (-1.5,5.75);
\draw [-](-1.5,5.75)  -- (-1,5.50);

\draw [-](-1.5,4.00)  -- (-2,4.00);
\draw [-](-1.5,5.00)  -- (-2,5.00);
\node [color=black] at (-2.5,4.00) {$0$};
\node [color=black] at (-2.5,5.00) {$\boldsymbol{\beta}_0$};

\draw [fill] (-1.75,5.00) circle [radius=.1];
\draw [-](-1.75,5.00)  -- (-1.75,10.00);
\draw [-](-1.75,10.00)  -- (0.00,10.00);

\end{tikzpicture}
  \caption{Proposed Fast-SSC datapath architecture to support operations from Table~\ref{tab:fast-instr} and Table~\ref{tab:instr:new}.}
  \label{fig:fasterssc}
\end{figure}
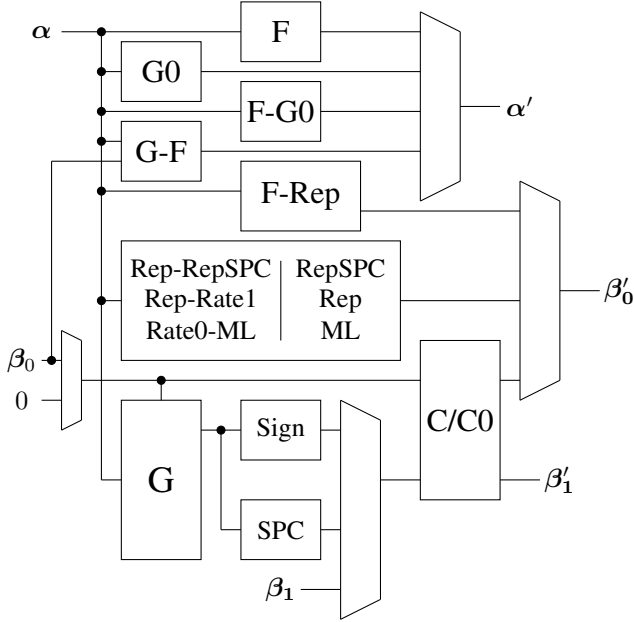

\subsection{Datapath}

Fig.~\ref{fig:fasterssc} shows the datapath architecture for the proposed Fast-SSC decoder, based on the implementation in \cite{sarkis14}. It supports all the operations listed in Table~\ref{tab:fast-instr} and Table~\ref{tab:instr:new}. The original critical path (in both Fig.~\ref{fig:fastssc} and Fig.~\ref{fig:fasterssc}) lies on the path G-SPC-C. To support multiple operations in a single cycle, the new G0 and C modules require more hardware complexity than the original G and C modules. Consequently, to avoid lengthening the critical path, modules including G and C operations are instantiated separately, and the old G and C modules are only used for P- prefixed operations from Table I. Modules F-G0 and G-F, and F-Rep are used only on the low-stage. Processing units for special nodes and their merged versions are clustered together in Fig~\ref{fig:fasterssc}; they receive LLR values from the $\alpha$ memory and output their hard decision estimates to $\beta$ memory.

\begin{figure}
  \centering
  \begin{tikzpicture}[scale=.55]

\draw (0.00,0.00) rectangle (1.00,15.00);
\draw [-](0.0,3.0)  -- (1.0,3.0);
\draw [-](0.0,6.0)  -- (1.0,6.0);
\draw [-](0.0,9.0)  -- (1.0,9.0);
\draw [-](0.0,12.0)  -- (1.0,12.0);

\draw [<->,thick](-0.25,0.0)  -- (-0.25,3.0);
\draw [<->,thick](-0.25,3.0)  -- (-0.25,6.0);
\draw [<->,thick](-0.25,6.0)  -- (-0.25,9.0);
\draw [<->,thick](-0.25,9.0)  -- (-0.25,12.0);
\draw [<->,thick](-0.25,12.0)  -- (-0.25,15.0);

\node [color=black] at (-0.75,13.50) {$P_e$};
\node [color=black] at (-0.75,10.50) {\Large $\frac{P_e}{2}$};
\node [color=black] at (-0.75,7.50) {\Large $\frac{P_e}{4}$};
\node [color=black] at (-0.75,4.50) {\Large $\frac{P_e}{8}$};
\node [color=black] at (-0.75,1.50) {\Large $\frac{P_e}{8}$};

\draw (10.00,0.00) rectangle (11.00,15.00);
\draw [-](10.0,3.0)  -- (11.0,3.0);
\draw [-](10.0,6.0)  -- (11.0,6.0);
\draw [-](10.0,9.0)  -- (11.0,9.0);
\draw [-](10.0,12.0)  -- (11.0,12.0);

\draw [<->,thick](11.25,0.0)  -- (11.25,3.0);
\draw [<->,thick](11.25,3.0)  -- (11.25,6.0);
\draw [<->,thick](11.25,6.0)  -- (11.25,9.0);
\draw [<->,thick](11.25,9.0)  -- (11.25,12.0);
\draw [<->,thick](11.25,12.0)  -- (11.25,15.0);

\node [color=black] at (11.75,13.50) {\Large $\frac{P_e}{2}$};
\node [color=black] at (11.75,10.50) {\Large $\frac{P_e}{4}$};
\node [color=black] at (11.75,7.50) {\Large $\frac{P_e}{8}$};
\node [color=black] at (11.75,4.50) {\Large $\frac{P_e}{16}$};
\node [color=black] at (11.75,1.50) {\Large $\frac{P_e}{16}$};

\draw (4.00,0.50) rectangle (7.00,2.50);
\draw (4.00,3.50) rectangle (7.00,5.50);
\draw (4.00,6.50) rectangle (7.00,8.50);
\draw (4.00,9.50) rectangle (7.00,11.50);
\draw (4.00,12.50) rectangle (7.00,14.50);

\node [color=black] at (5.50,13.50) {\Large $\text{F}_{\frac{P_e}{2}}$};
\node [color=black] at (5.50,10.50) {\Large $\text{F}_{\frac{P_e}{4}}$};
\node [color=black] at (5.50,7.50) {\Large $\text{F}_{\frac{P_e}{8}}$};
\node [color=black] at (5.50,4.50) {\Large $\text{F}_{\frac{P_e}{16}}$};
\node [color=black] at (5.50,1.50) {\Large $\text{F}_{\frac{P_e}{16}}$};

\draw [->](7.0,1.5)  -- (10.0,1.5);
\draw [->](7.0,4.5)  -- (10.0,4.5);
\draw [->](7.0,7.5)  -- (10.0,7.5);
\draw [->](7.0,10.5)  -- (10.0,10.5);
\draw [->](7.0,13.5)  -- (10.0,13.5);

\draw [fill] (8.5,13.5) circle [radius=.1];
\draw [fill] (8.5,10.5) circle [radius=.1];
\draw [fill] (8.5,7.5) circle [radius=.1];

\draw [-](2.5,3.5)  -- (2.5,5.5);
\draw [-](2.5,5.5)  -- (3.0,5.0);
\draw [-](3.0,5.0)  -- (3.0,4.0);
\draw [-](3.0,4.0)  -- (2.5,3.5);

\draw [-](2.5,6.5)  -- (2.5,8.5);
\draw [-](2.5,8.5)  -- (3.0,8.0);
\draw [-](3.0,8.0)  -- (3.0,7.0);
\draw [-](3.0,7.0)  -- (2.5,6.5);

\draw [-](2.5,9.5)  -- (2.5,11.5);
\draw [-](2.5,11.5)  -- (3.0,11.0);
\draw [-](3.0,11.0)  -- (3.0,10.0);
\draw [-](3.0,10.0)  -- (2.5,9.5);

\draw [->](1.0,1.5)  -- (4.0,1.5);
\draw [->](1.0,13.5)  -- (4.0,13.5);

\draw [-](8.5,13.5)  -- (8.5,12.0);
\draw [-](8.5,12.0)  -- (2.0,12.0);
\draw [-](2.0,12.0)  -- (2.0,11.00);
\draw [->](2.0,11.00)  -- (2.5,11.00);
\draw [->](1.0,10.0)  -- (2.5,10.00);
\draw [->](3.0,10.5)  -- (4.0,10.5);

\draw [-](8.5,10.5)  -- (8.5,9.0);
\draw [-](8.5,9.0)  -- (2.0,9.0);
\draw [-](2.0,9.0)  -- (2.0,8.00);
\draw [->](2.0,8.00)  -- (2.5,8.00);
\draw [->](1.0,7.0)  -- (2.5,7.00);
\draw [->](3.0,7.5)  -- (4.0,7.5);

\draw [-](8.5,7.5)  -- (8.5,6.0);
\draw [-](8.5,6.0)  -- (2.0,6.0);
\draw [-](2.0,6.0)  -- (2.0,5.00);
\draw [->](2.0,5.00)  -- (2.5,5.00);
\draw [->](1.0,4.0)  -- (2.5,4.00);
\draw [->](3.0,4.5)  -- (4.0,4.5);

\end{tikzpicture}
  \caption{Reconfigurable processing unit for F branch operations. The architecture can support multiple F operations at low-stage, as well as single F operations at high-stage.}
  \label{fig:arch_mergedF}
\end{figure}
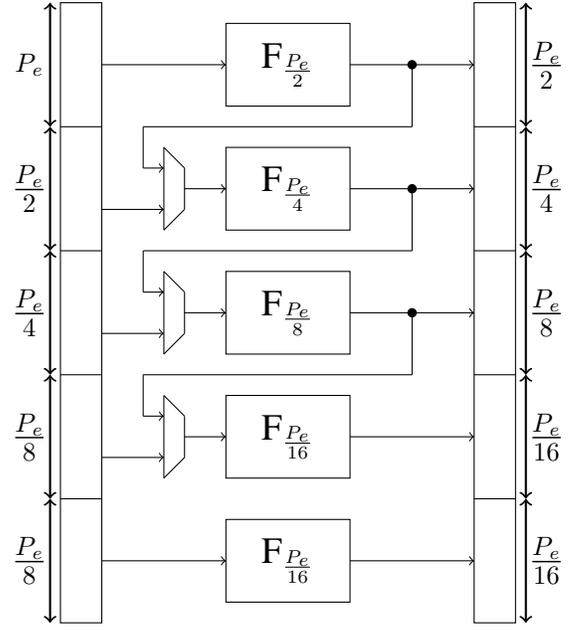

In order to support single operations of high-stage as well as single and multiple operations of low-stage, the F processing unit has been redesigned and is shown in Fig.~\ref{fig:arch_mergedF}. The new F processing unit is capable of performing high-stage operations in multiple clock cycles, as well as performing single and multiple operations at low-stage. The subscripts in Fig.~\ref{fig:arch_mergedF} correspond to the number of parallel F processing elements, which is $P_e$ in total. At high-stage operations, the inputs of all F processing units in Fig.~\ref{fig:arch_mergedF} are provided from the $\alpha$ memory. For multiple operations at low-stage, the multiplexers are configured by the controller to cascade the processing units. The configuration of multiple operations are established through the multiplexers in the design: if a merged F operation is performed, following operations within the merged F operation take their inputs from the output of a previous F module. Similar architectures have been implemented for G0 and C/C0 modules.

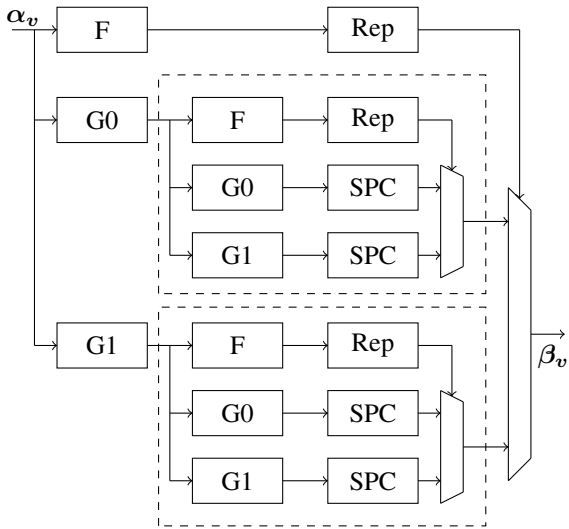
\begin{figure}[]
  \centering
  \scalebox{1.00}{
  \begin{tikzpicture}[scale=.60]

\draw (-3.0,3.0) rectangle (-1.0,4.0);
\node [color=black] at (-2.0,3.50) {G1};

\draw (0.0,3.0) rectangle (2.0,4.0);
\node [color=black] at (1.0,3.50) {F};
\draw (0.0,1.5) rectangle (2.0,2.5);
\node [color=black] at (1.0,2.00) {G0};
\draw (0.0,0.0) rectangle (2.0,1.0);
\node [color=black] at (1.0,0.50) {G1};

\draw (3.0,3.0) rectangle (5.0,4.0);
\node [color=black] at (4.0,3.50) {Rep};
\draw (3.0,1.5) rectangle (5.0,2.5);
\node [color=black] at (4.0,2.00) {SPC};
\draw (3.0,0.0) rectangle (5.0,1.0);
\node [color=black] at (4.0,0.50) {SPC};

\draw [->](-1.0,3.5)  -- (0.00,3.5);
\draw [-](-0.5,3.5)  -- (-0.5,0.5);
\draw [->](-0.5,2.0)  -- (-0.0,2.0);
\draw [->](-0.5,0.5)  -- (-0.0,0.5);

\draw [->](2.0,0.5)  -- (3.00,0.5);
\draw [->](2.0,2.0)  -- (3.00,2.0);
\draw [->](2.0,3.5)  -- (3.00,3.5);

\draw [dashed] (-0.75,-0.5) rectangle (6.5,4.35);

\draw [->](5.0,0.5)  -- (5.50,0.5);
\draw [->](5.0,2.0)  -- (5.50,2.0);
\draw [-] (5.0,3.5)  -- (5.75,3.5); 

\draw [-] (5.5,0.0)  -- (5.50,2.5);
\draw [-] (5.50,2.5)  -- (6.00,2.25);
\draw [-] (6.00,2.25)  -- (6.00,0.25);
\draw [-] (6.00,0.25)  -- (5.5,0.0);
\draw [->] (5.75,3.5)  -- (5.75,2.375); 

\draw [->] (6.00,1.25)  -- (7.00,1.25);


\draw (-3.0,8.0) rectangle (-1.0,9.0);
\node [color=black] at (-2.0,8.50) {G0};

\draw (0.0,8.0) rectangle (2.0,9.0);
\node [color=black] at (1.0,8.50) {F};
\draw (0.0,6.5) rectangle (2.0,7.5);
\node [color=black] at (1.0,7.00) {G0};
\draw (0.0,5.0) rectangle (2.0,6.0);
\node [color=black] at (1.0,5.50) {G1};

\draw (3.0,8.0) rectangle (5.0,9.0);
\node [color=black] at (4.0,8.50) {Rep};
\draw (3.0,6.5) rectangle (5.0,7.5);
\node [color=black] at (4.0,7.00) {SPC};
\draw (3.0,5.0) rectangle (5.0,6.0);
\node [color=black] at (4.0,5.50) {SPC};

\draw [->](-1.0,8.5)  -- (0.00,8.5);
\draw [-](-0.5,8.5)  -- (-0.5,5.5);
\draw [->](-0.5,7.0)  -- (-0.0,7.0);
\draw [->](-0.5,5.5)  -- (-0.0,5.5);

\draw [->](2.0,5.5)  -- (3.00,5.5);
\draw [->](2.0,7.0)  -- (3.00,7.0);
\draw [->](2.0,8.5)  -- (3.00,8.5);

\draw [dashed] (-0.75,4.65) rectangle (6.5,9.5);

\draw [->](5.0,5.5)  -- (5.50,5.5);
\draw [->](5.0,7.0)  -- (5.50,7.0);
\draw [-] (5.0,8.5)  -- (5.75,8.5); 

\draw [-] (5.5,5.0)  -- (5.50,7.5);
\draw [-] (5.50,7.5)  -- (6.00,7.25);
\draw [-] (6.00,7.25)  -- (6.00,5.25);
\draw [-] (6.00,5.25)  -- (5.5,5.0);
\draw [->] (5.75,8.5)  -- (5.75,7.375); 

\draw [->] (6.00,6.25)  -- (7.00,6.25);


\draw (-3.0,10.0) rectangle (-1.0,11.0);
\node [color=black] at (-2.0,10.50) {F};

\draw [->] (-1.0,10.5)  -- (3.00,10.5);

\draw (3.0,10.0) rectangle (5.0,11.0);
\node [color=black] at (4.0,10.50) {Rep};

\draw [-] (5.00,10.5)  -- (7.25,10.5);
\draw [->] (7.25,10.5)  -- (7.25,6.75);

\draw [-] (7.00,7.00)  -- (7.00,0.50);
\draw [-] (7.00,0.50)  -- (7.50,1.00);
\draw [-] (7.50,1.00)  -- (7.50,6.50);
\draw [-] (7.50,6.50)  -- (7.00,7.00);


\node [color=black] at (-3.75,10.75) {$\boldsymbol{\alpha_v}$};

\draw [->] (-4.0,10.5)  -- (-3.00,10.5);
\draw [-] (-3.5,10.5)  -- (-3.50,3.5);
\draw [->] (-3.50,3.5)  -- (-3.00,3.5);
\draw [->] (-3.50,8.5)  -- (-3.00,8.5);

\draw [->] (7.50,3.75)  -- (8.25,3.75);
\node [color=black] at (8.00,3.25) {$\boldsymbol{\beta_v}$};

\end{tikzpicture}}
  \caption{Architecture of Rep-RepSPC processor.}
  \label{fig:arch:RepRepSPC}
\end{figure}

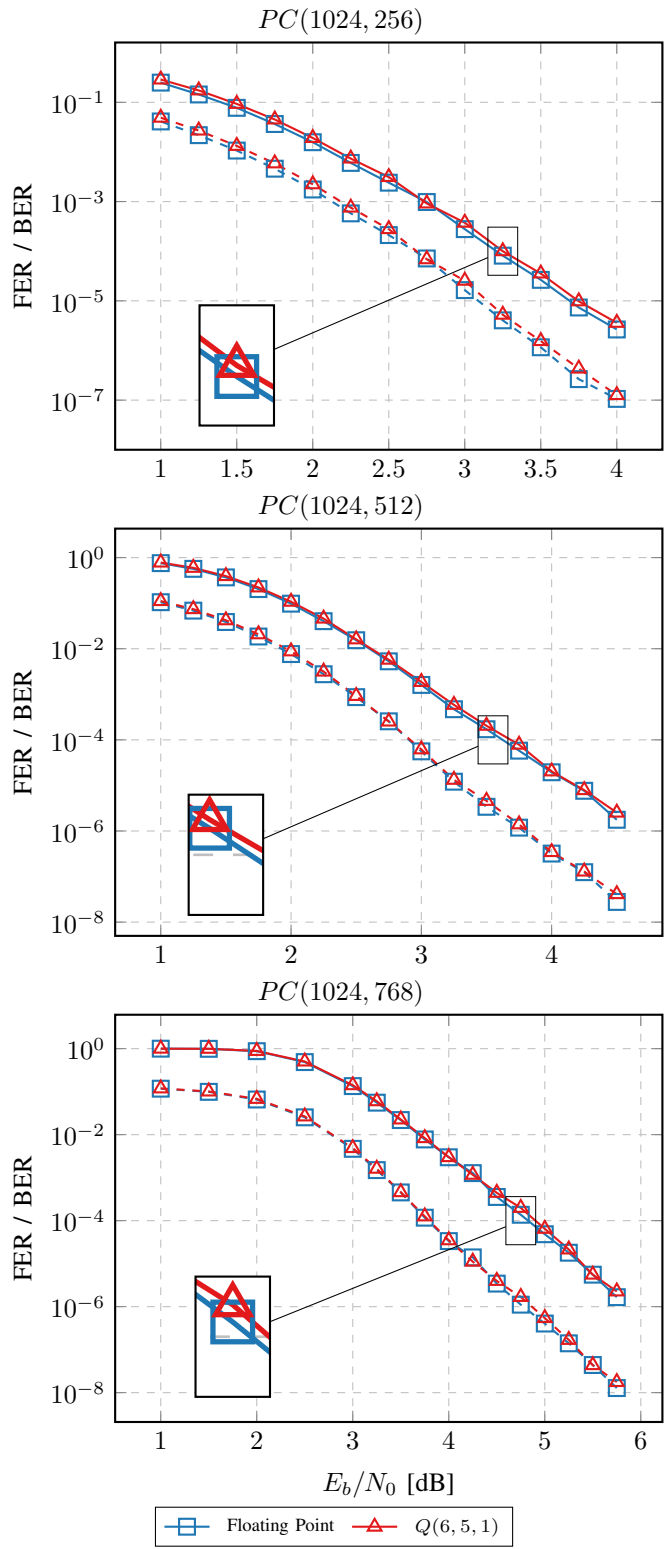
\begin{figure}
\centering
\begin{tikzpicture}[spy using outlines=
  {rectangle, magnification=2.5, connect spies}]

\begin{axis}[
  scale = 1,
    ymode=log,
    ylabel={FER~/~BER},
    grid=both,
    ymajorgrids=true,
    xmajorgrids=true,
    ymin = 0.00000001,
    grid style=dashed,
    width=\columnwidth, height=7cm,
    thick,
    mark size=3,
]

\addplot[
    color=Paired-1,
    mark=square,
    thick,
    mark size=3,
]
table {
1.0 2.49332e-01
1.25  1.43858e-01
1.5 7.73779e-02
1.75  3.65304e-02
2.0 1.57029e-02
2.25  6.01196e-03
2.5 2.40279e-03
2.75  9.79432e-04
3.0 2.78552e-04
3.25  8.14067e-05
3.5 2.64725e-05
3.75  7.35078e-06
4.0 2.65076e-06
};

\addplot[
    color=Paired-5,
    mark=triangle,
    thick,
    mark size=3,
]
table {
1.0 2.87737e-01
1.25  1.73599e-01
1.5 9.28116e-02
1.75  4.50947e-02
2.0 1.91027e-02
2.25  7.35793e-03
2.5 3.12064e-03
2.75  9.16590e-04
3.0 3.75235e-04
3.25  1.00060e-04
3.5 3.55619e-05
3.75  9.78521e-06
4.0 3.60719e-06
};

\addplot[
    color=Paired-1,
    dashed,
    mark=square,
    mark options={solid},
    thick,
    mark size=3,
]
table {
1.0 4.11403e-02
1.25  2.17823e-02
1.5 1.06580e-02
1.75  4.59874e-03
2.0 1.75224e-03
2.25  5.78849e-04
2.5 2.11319e-04
2.75  7.14297e-05
3.0 1.64628e-05
3.25  4.05126e-06
3.5 1.16024e-06
3.75  2.64743e-07
4.0 1.05823e-07
};

\addplot[
    color=Paired-5,
    dashed,
    mark=triangle,
    mark options={solid},
    thick,
    mark size=3,
]
table {
1.0 4.89273e-02
1.25  2.69796e-02
1.5 1.32302e-02
1.75  5.88418e-03
2.0 2.23085e-03
2.25  7.56130e-04
2.5 2.78617e-04
2.75  6.88159e-05
3.0 2.54163e-05
3.25  5.25315e-06
3.5 1.54750e-06
3.75  4.31543e-07
4.0 1.26252e-07
};

\coordinate (spypoint) at (axis cs:3.25,1e-4);
\coordinate (magnifyglass) at (axis cs:1.5,5e-7);

\end{axis}
\spy [black, height=1.6cm, width=.99cm] on (spypoint)
   in node[fill=white] at (magnifyglass);

\node [color=black] at (3.0,5.7) {$PC(1024,256)$} ;
\end{tikzpicture}
\\
\begin{tikzpicture}[spy using outlines=
  {rectangle, magnification=2.5, connect spies}]
\begin{axis}[
  scale = 1,
    ymode=log,
    ylabel={FER~/~BER},
    grid=both,
    ymajorgrids=true,
    xmajorgrids=true,
    grid style=dashed,
    width=\columnwidth, height=7cm,
    thick,
    mark size=3,
    legend style={
      anchor={center},
      cells={anchor=west},
      column sep= 1.5mm,
      font=\fontsize{7pt}{7.2}\selectfont,
    },
    legend to name=perf-ber,
    legend columns=2,
]

\addplot[
    color=Paired-1,
    mark=square,
    thick,
    mark size=3,
]
table {
1.0 7.56730e-01
1.25  5.66600e-01
1.5 3.68275e-01
1.75  2.03978e-01
2.0 9.75773e-02
2.25  4.02692e-02
2.5 1.53739e-02
2.75  5.29412e-03
3.0 1.60518e-03
3.25  4.70146e-04
3.5 1.71028e-04
3.75  5.79643e-05
4.0 1.94382e-05
4.25  7.62091e-06
4.5 1.76080e-06
};
\addlegendentry{Floating Point}

\addplot[
    color=Paired-5,
    mark=triangle,
    thick,
    mark size=3,
]
table {
1.0 7.79212e-01 
1.25  5.96570e-01
1.5 3.93240e-01
1.75  2.21705e-01
2.0 1.07507e-01
2.25  4.58425e-02
2.5 1.60518e-02
2.75  5.74277e-03
3.0 1.86441e-03
3.25  5.87199e-04
3.5 2.03832e-04
3.75  7.83269e-05
4.0 2.03182e-05
4.25  7.87879e-06
4.5 2.52682e-06
};
\addlegendentry{$Q(6,5,1)$}

\addplot[
    color=Paired-1,
    dashed,
    mark=square,
    mark options={solid},
    thick,
    mark size=3,
]
table {
1.0 1.05130e-01
1.25  6.86705e-02
1.5 3.86167e-02
1.75  1.85484e-02
2.0 7.64556e-03
2.25  2.75605e-03
2.5 8.66210e-04
2.75  2.55269e-04
3.0 5.59260e-05
3.25  1.20750e-05
3.5 3.39383e-06
3.75  1.18419e-06
4.0 3.19288e-07
4.25  1.24733e-07
4.5 2.77189e-08
};

\addplot[
    color=Paired-5,
    dashed,
    mark=triangle,
    thick,
    mark size=3,
    mark options={solid}
]
table {
1.0 1.11173e-01 
1.25  7.39820e-02
1.5 4.21431e-02
1.75  2.07406e-02
2.0 8.75619e-03
2.25  3.14954e-03
2.5 9.19916e-04
2.75  2.52348e-04
3.0 6.18457e-05
3.25  1.32808e-05
3.5 4.57826e-06
3.75  1.39826e-06
4.0 3.44457e-07
4.25  1.28646e-07
4.5 4.10608e-08
};

\coordinate (spypoint) at (axis cs:3.55,1e-4);
\coordinate (magnifyglass) at (axis cs:1.5,3e-7);

\end{axis}
\spy [black, height=1.6cm, width=.99cm] on (spypoint)
   in node[fill=white] at (magnifyglass);

\node [color=black] at (3.0,5.7) {$PC(1024,512)$} ;
\end{tikzpicture}
\begin{tikzpicture}[spy using outlines=
  {rectangle, magnification=2.5, connect spies}]
\begin{axis}[
  scale = 1,
    ymode=log,
    xlabel={$E_b/N_0$ [\text{dB}]}, xlabel style={yshift=0em},
    ylabel={FER~/~BER},
    grid=both,
    ymajorgrids=true,
    xmajorgrids=true,
    grid style=dashed,
    width=\columnwidth, height=7cm,
    thick,
    mark size=3,
]

\addplot[
    color=Paired-1,
    mark=square,
    thick,
    mark size=3,
]
table {
1.0 9.99860e-01
1.5 9.91755e-01
2.0 8.74337e-01
2.5 4.87637e-01
3.0 1.34826e-01
3.25  5.55234e-02
3.5 2.18644e-02
3.75  7.74676e-03
4.0 2.98106e-03
4.25  1.27617e-03
4.5 3.55492e-04
4.75  1.37155e-04
5.0 4.83699e-05
5.25  1.79814e-05
5.5 5.54723e-06
5.75  1.66317e-06
};

\addplot[
    color=Paired-5,
    mark=triangle,
    thick,
    mark size=3,
]
table {
1.0 9.99801e-01
1.5 9.92911e-01
2.0 8.85723e-01
2.5 5.02881e-01
3.0 1.41406e-01
3.25  5.99402e-02
3.5 2.25922e-02
3.75  8.30508e-03
4.0 3.00100e-03
4.25  1.16650e-03
4.5 4.43853e-04
4.75  1.98295e-04
5.0 6.46663e-05
5.25  2.13945e-05
5.5 5.55960e-06
5.75  2.26087e-06
};

\addplot[
    color=Paired-1,
    dashed,
    mark=square,
    thick,
    mark options={solid},
    mark size=3,
]
table {
1.0 1.17354e-01
1.5 9.90862e-02
2.0 6.57489e-02
2.5 2.53016e-02
3.0 4.68549e-03
3.25  1.49902e-03
3.5 4.54067e-04
3.75  1.17356e-04
4.0 3.34803e-05 
4.25  1.39945e-05
4.5 3.43920e-06
4.75  1.11260e-06
5.0 4.04972e-07
5.25  1.40480e-07
5.5 4.39878e-08
5.75  1.27986e-08
};

\addplot[
    color=Paired-5,
    dashed,
    mark=triangle,
    thick,
    mark options={solid},
    mark size=3,
]
table {
1.0 1.19024e-01
1.5 1.01005e-01
2.0 6.79553e-02
2.5 2.65637e-02
3.0 4.87498e-03
3.25  1.61332e-03
3.5 4.59637e-04
3.75  1.25736e-04
4.0 3.49927e-05
4.25  1.11385e-05
4.5 3.97040e-06
4.75  1.68086e-06
5.0 5.43096e-07
5.25  1.68816e-07
5.5 4.36516e-08
5.75  1.76631e-08
};

\coordinate (spypoint) at (axis cs:4.75,1e-4);
\coordinate (magnifyglass) at (axis cs:1.75,2e-7);

\end{axis}
\spy [black, height=1.6cm, width=.99cm] on (spypoint)
   in node[fill=white] at (magnifyglass);

\node [color=black] at (3.0,5.7) {$PC(1024,768)$} ;
\end{tikzpicture}
\\
\ref{perf-ber}
\caption{Floating-point and quantized FER and BER performance for polar code with $N=1024$ and $R \in \{\frac{1}{3},\frac{1}{2},\frac{2}{3}\}$. Polar code construction is obtained  from the 5G standard  \cite{38.212}. Solid and dashed lines represent FER and BER performances respectively.}
\label{fig:BER}
\end{figure}

Fig.~\ref{fig:arch:RepRepSPC} presents the Rep-RepSPC processing unit from Table~\ref{tab:instr:new}. The input and the output size of the Rep-RepSPC has a fixed length of $16$. The units enclosed with dashed lines are instances of the RepSPC processing unit. The RepSPC modules assume the output of the Rep node as $0$ and $1$ and are processed in parallel with the Rep module. Modules G0 and G1 are G operations that assumes $\beta$ as $0$ and $1$, respectively. Based on the hard decision estimate of Rep node, the output of the RepSPC is selected from the final multiplexer and is stored in the $\beta$ memory. 

\section{Results}\label{sec:results}

\begin{table*}[h]
\centering
\caption{TSMC 65 nm CMOS implementation results for Fast-SSC based polar decoders, $N=1024$, $R \in \{\frac{1}{4},\frac{1}{2},\frac{3}{4}\}$ and $P_e = 64$. }
\label{tab:asicresults-fastssc}
\resizebox*{1.6\columnwidth}{!}{
\hspace*{-6em}
\begin{tabular}{@{}lllllllllllll@{}}
\toprule
                     & \multicolumn{3}{c}{\cite{sarkis14}} & \multicolumn{3}{c}{\cite{giardJSPS}} & \multicolumn{3}{c}{\cite{fastssc-sips17}} & \multicolumn{3}{c}{This Work} \\
                       \cmidrule(l){2-4}               \cmidrule(l){5-7}               \cmidrule(l){8-10}              \cmidrule(l){11-13}
Algorithm            & \multicolumn{3}{c}{Fast-SSC}  & \multicolumn{3}{c}{Fast-SSC}  & \multicolumn{3}{c}{Fast-SSC}  & \multicolumn{3}{c}{Fast-SSC}  \\
$P_e$                & \multicolumn{3}{c}{64}        & \multicolumn{3}{c}{64}        & \multicolumn{3}{c}{64}        & \multicolumn{3}{c}{64}        \\
Technology (nm)      & \multicolumn{3}{c}{65}        & \multicolumn{3}{c}{65}        & \multicolumn{3}{c}{65}        & \multicolumn{3}{c}{65}        \\
Quantization         & \multicolumn{3}{c}{Q(6,5,1)}  & \multicolumn{3}{c}{Q(6,5,1)}  & \multicolumn{3}{c}{Q(6,5,1)}  & \multicolumn{3}{c}{Q(6,5,1)}  \\
Supply(V)               & \multicolumn{3}{c}{1.0}       & \multicolumn{3}{c}{1.0}       & \multicolumn{3}{c}{1.0}       & \multicolumn{3}{c}{1.0}       \\
Power (mW)           & \multicolumn{3}{c}{160.78}    & \multicolumn{3}{c}{114.00}    & \multicolumn{3}{c}{173.75}    & \multicolumn{3}{c}{189.09}          \\

Area (mm$^2$)        & \multicolumn{3}{c}{0.60}    & \multicolumn{3}{c}{0.44}    & \multicolumn{3}{c}{0.57}    & \multicolumn{3}{c}{0.64}          \\

Frequency (MHz)      & \multicolumn{3}{c}{450}       & \multicolumn{3}{c}{450}       & \multicolumn{3}{c}{420}       & \multicolumn{3}{c}{430}          \\
                       \cmidrule(l){2-4}               \cmidrule(l){5-7}               \cmidrule(l){8-10}              \cmidrule(l){11-13}
Rate                 & 1/4      & 1/2      & 3/4     & 1/4      & 1/2      & 3/4     & 1/4      & 1/2      & 3/4     & 1/4      & 1/2      & 3/4     \\
\cmidrule(l){2-2}\cmidrule(l){3-3}\cmidrule(l){4-4} \cmidrule(l){5-5}\cmidrule(l){6-6}\cmidrule(l){7-7}\cmidrule(l){8-8}\cmidrule(l){9-9}\cmidrule(l){10-10}\cmidrule(l){11-11}\cmidrule(l){12-12}\cmidrule(l){13-13}
Latency ($\mu$s)    & 0.5      & 0.6      & 0.5     & 0.49     & 0.56     & 0.48    & 0.45     & 0.51     & 0.44    &    0.39   &   0.46   &  0.42   \\
Coded T/P (Mpbs)           & 2030     & 1719     & 2039    & 2104     & 1829     & 2114    & 2288     & 2000     & 2337    &    2653   &   2213   &  2433   \\
Info. T/P (Mpbs)     & 507      & 860      & 1529    & 526      & 914      & 1585    & 572      & 1000     & 1753    &    663    &   1106   &  1825   \\
Area Eff. (Gbps/mm$^2$) & 3.38     & 2.87     & 3.4     & 4.78     & 4.16     & 4.8     & 4.01     & 3.51     & 4.10    &    4.14   &   3.46   &  3.80   \\
Energy (pJ/bit)      & 316.81   & 187.02   & 105.14  & 216.72   & 124.69   & 71.91   & 303.80   & 173.72   & 99.11   &    285.15 &   170.92 &  103.64 \\
\bottomrule
\end{tabular}}
\end{table*}

\subsection{Error-Correction Performance}

To validate the error-correction performance for the proposed decoder, a quantization scheme $Q(6,5,1)$ has been used, where $Q(Q_i,Q_c,Q_f)$ are quantization bit size for internal LLRs, channel LLRs, and fraction bit size for both internal and channel LLRs, respectively. Fig.~\ref{fig:BER} depicts the error correction performance of the proposed decoder in terms of bit error rate (BER) and frame error rate (FER). The polar code construction is obtained from \cite{38.212} for $N=1024$, and rates are selected as $R \in \{\frac{1}{4},\frac{1}{2},\frac{3}{4}\}$. The selected quantization values result in less than $0.03$ dB loss at FER$=10^{-4}$ compared to floating-point precision. The introduced operation merging techniques do not change the error-correction performance of SC decoding, as they map thoroughly to SC decoding schedule.

\subsection{ASIC Synthesis Results}

The architecture for the proposed Fast-SSC decoder has been implemented in VHDL and synthesized in TSMC 65 nm CMOS technology using Cadence Genus RTL compiler. For a fair comparison scheme, three other Fast-SSC-based decoders from \cite{sarkis14, giardJSPS, fastssc-sips17} have also been implemented using the same technology node, quantization, voltage supply and $P_e$. Table~\ref{tab:asicresults-fastssc} presents the ASIC implementation results for code rates $R \in \{\frac{1}{4},\frac{1}{2},\frac{3}{4}\}$. 

According to Table~\ref{tab:asicresults-fastssc}, the proposed Fast-SSC decoder has a throughput improvement of up to $31\%$ and $26\%$ compared to the earlier implementations from \cite{sarkis14} and \cite{giardJSPS}, respectively.  Compared to our previous work in \cite{fastssc-sips17}, the throughput improvement is up to $16\%$. The power consumption of the proposed decoder has increased by $18\%$ compared to the baseline Fast-SSC decoder from \cite{sarkis14}, which is due to the new decoding nodes introduced in Section~\ref{sec:architecture}. On the other hand, due to increased throughput, energy consumption and area efficiency of the proposed decoder compared to \cite{sarkis14} has been improved by up to $10\%$ and $23\%$ despite the increased power consumption. On the other hand, compared to \cite{giardJSPS}, the proposed decoder implementation consumes $32\%$ more energy per bit and has $13\%$ less area efficiency. 

It can be observed in Table~\ref{tab:asicresults-fastssc} that, for all Fast-SSC-based implementations, the latency and the coded throughput are the lowest when $R = 1/2$. This is due to the fact that the occurrence of Rate-0 nodes increase when rate becomes lower, and Rate-1 nodes increase when rate becomes higher. Around $R \approx 1/2$, Rep, SPC and ML nodes occur more frequently, which in general takes more time for decoding. As a result, the area efficiency has the same trend with latency and coded throughput with respect to code rate. On the other hand, the information throughput increases with the rate for all Fast-SSC-based implementations since the number of information bits in the codeword increases linearly with the rate. Finally, the energy dissipation per decoded bit is calculated using the number of information bits. As a result, energy per bit reduces with increasing rate for all Fast-SSC decoders in Table~\ref{tab:asicresults-fastssc}.

\begin{table*}[]
\centering
\caption{Comparison of state-of-the-art ASIC implementations decoding a $PC(1024,512)$ polar code.}
\label{tab:asicresults-others}
\resizebox*{1.6\columnwidth}{!}{%
\hspace{-6em}
\begin{tabular}{@{}llllllll@{}}
\toprule
                              & This Work & \cite{redundantSC} & \cite{SPSC13} & \cite{SC_2bit} & \cite{Mishra2012} & \cite{dizdar16} & \cite{SC-Fan14} \\
                              \cmidrule(l){2-2} \cmidrule(l){3-3} \cmidrule(l){4-4} \cmidrule(l){5-5} \cmidrule(l){6-6} \cmidrule(l){7-7} \cmidrule(l){8-8}
Algorithm                     & Fast-SSC  & Tree                 & Semi-Parallel   & Tree              & Semi-Parallel       & Combinational     & Semi-Parallel     \\
$P_e$               & 64        & -                  & 64              & -                & 64                 & -                & 64                \\
Technology (nm)               & 65        & 180                  & 65              & 45                & 180                 & 90                & 65                \\
Supply (V)                    & 1.0       & -                    & 1.0             & -                 & 1.3                 & 1.3               & 1.2               \\
Area (mm$^2$)                 & 0.64      & -                    & 0.31            & -                 & 1.71                & 3.21              & 0.68              \\
Area @ 65 nm (mm$^2$)         & 0.64      & -                    & 0.31            & -                 & 0.22                & 1.68              & 0.68              \\
Frequency (MHz)               & 430       & 377                  & 500             & 750               & 150                 & 2.5               & 1010              \\
Throughput (Mbps) \textsuperscript{\textdagger}          & 1106      & 349                  & 123             & 346               & 136                 & 3544              & 497               \\
Area Efficiency (Gbps/mm$^2$) \textsuperscript{\textdagger} & 3.46      & -                    & 0.39            & -                 & 0.62                & 2.11              & 0.73 \\            
\bottomrule
\textdagger \scriptsize Scaled for 65 nm technology.
\end{tabular}}
\end{table*}

Table~\ref{tab:asicresults-others} presents a comparison scheme for the proposed Fast-SSC decoder against other SC-based architectures including tree, semi-parallel (SP) and combinational approaches. The throughput values for each implementation in Table~\ref{tab:asicresults-others} are scaled for 65 nm for a fair comparison. 
In \cite{dizdar16}, a combinational approach is used to decode the polar code, which results in low operating frequency, large throughput and increased area. Although the reported throughput is $2.2 \times$ higher than our decoder, the area is $2.62 \times$ larger, which results in $39\%$ less area efficiency compared to the proposed architecture. In fact, compared with any Fast-SSC based architecture from Table~\ref{tab:asicresults-fastssc}, it can be observed that the area efficiency of the combinational decoder is very low due to its excessive area overhead. Compared to semi-parallel decoder implementations with equal $P_e$, our work has up to $9\times$ larger throughput and $8.8\times$ better area efficiency. Finally, compared with tree-based decoder approaches, the proposed decoder has $3.18\times$ larger throughput.

\section{Conclusion}\label{sec:conclusion}
In this work, we proposed a new Fast-SSC polar code decoder implementation. The proposed decoder increases the memory utilization by storing the variables relative to the stages below the parallelization threshold into a single memory word, which also enables the decoder to perform multiple operations within a single time step. A generalization of operation merging scenarios with their guidelines are presented for branch and special node operations, and a subset of the merging scenarios are selected to be implemented in hardware. With the proposed technique, the memory utilization is risen from $66\%$ to $99.6\%$, while the proposed operation set reduces the number of operations to decode a codeword by up to $35\%$ compared to the baseline Fast-SSC decoder. Proposed decoder has been implemented in TSMC 65 nm technology node and compared against other Fast-SSC-based implementations. Results show that our decoder implementation has a throughput improvement of up to $31\%$ and $26\%$ compared to the earlier Fast-SSC-based decoder implementations, with a slight increase in power consumption. Energy dissipation per decoded information bit and the area efficiency for the proposed decoder has been improved by $10\%$ and $23\%$ compared to the baseline Fast-SSC decoder. Compared to semi-parallel and tree decoder implementations, proposed decoder has an up to $9\times$ larger throughput and $8.8\times$ better area efficiency.

\end{document}